\documentclass[final,5p,times,twocolumn]{elsarticle}
\usepackage{algorithm}
\usepackage{algpseudocode}
\usepackage{pifont}
\newcommand{\cmark}{\ding{51}}%
\newcommand{\xmark}{\ding{55}}%
\usepackage{setspace}
\usepackage{graphicx,natbib}

\usepackage{float, subcaption}
\usepackage{amssymb,amsmath,multicol, multirow}
\usepackage[english]{babel} 
\usepackage{siunitx}  
\usepackage[hidelinks]{hyperref}

\usepackage{cleveref}
\usepackage[font=normalsize]{caption}
\usepackage{url}

\usepackage[utf8]{inputenc} 
\usepackage[T1]{fontenc}    
\usepackage{booktabs}       
\usepackage{amsfonts}       
\usepackage{nicefrac}       
\usepackage{microtype}      %
\usepackage{fancyhdr}       
\usepackage{booktabs}
\usepackage{chngcntr}

\usepackage[monochrome]{color}
\usepackage[disable]{todonotes}

\usepackage{color}
\usepackage{todonotes}
\newcommand{\rtodo}[1]{\todo[color=red!80]{\scriptsize #1}} 

\usepackage{tcolorbox}

\journal{}
\date{26 October 2024}

\begin{document}

\begin{frontmatter}

\title{Interpretable ECG Analysis for Myocardial Infarction Detection through Counterfactuals}

\author[itu]{Toygar Tanyel}

\author[mae]{Sezgin Atmaca}
\author[mae]{Kaan Gökçe}
\author[aalto]{M. Yiğit Balık}
\author[mae]{Arda Güler}
\author[mu]{Emre Aslanger}
\author[itu2]{İlkay Öksüz\corref{cor1}}
\ead{oksuzilkay@itu.edu.tr}
\cortext[cor1]{Corresponding Author, \textit{Addressline:} Istanbul Technical University,
ITU Ayazaga Kampusu, Bilgisayar-Bilisim Fakultesi,
Maslak, 34467 Sariyer/Istanbul, \textit{Fax:} 0212 285 34 24, \textit{Tel:} 90-212-285-3593}

\affiliation[itu]{organization={Biomedical Engineering Graduate Program, Istanbul Technical University}, city={Istanbul}, country={Turkiye}}

\affiliation[mae]{organization={Mehmet Akif Ersoy Thoracic and Cardiovascular Surgery Training and Research Hospital}, city={Istanbul}, country={Turkiye}}

\affiliation[aalto]{organization={Department of Computer Science, Aalto University}, city={Espoo}, country={Finland}}

\affiliation[mu]{organization={Department of Cardiology, Başakşehir Pine and Sakura City Hospital}, city={Istanbul}, country={Turkiye}}

\affiliation[itu2]{organization={Department of Computer Engineering, Istanbul Technical University}, city={Istanbul}, country={Turkiye}}

\begin{abstract}

In the evolving landscape of ECG signal analysis, \textcolor{blue}{the challenge of limited transparency in machine learning models remains a significant barrier to their effective integration into clinical practice}. \textcolor{blue}{This study addresses this issue by investigating the use of counterfactual explanations to improve model interpretability for clinicians, particularly in differentiating healthy subjects from myocardial infarction patients.} Utilizing the PTB-XL dataset, we developed a methodology for systematic feature extraction and refinement to prepare for counterfactual analysis. This led to the creation of the Visualizing Counterfactual Clues on Electrocardiograms (VCCE) method, designed to improve the practicality of counterfactual explanations in a clinical setting. The validity of our approach was assessed using custom metrics that reflect the diagnostic relevance of counterfactuals, evaluated with the help of two cardiologists. Our findings suggest that this approach could support future efforts in using ECGs to predict patient outcomes for cardiac conditions, achieving interpretation validity scores of 23.29 ± 1.04 and 20.28 ± 0.99 out of 25 for high and moderate-quality interpretations, respectively. Clinical alignment scores of 0.83 ± 0.12 for high-quality and 0.57 ± 0.10 for moderate-quality interpretations underscore the potential clinical applicability of our method. The methodology and findings of this study contribute to the ongoing discussion on enhancing the interpretability of machine learning models in cardiology, offering a concept that bridges the gap between advanced data analysis techniques and clinical decision-making. The source code for this study is available at https://github.com/tanyelai/vcce.
\end{abstract}


\end{frontmatter}


\section{Introduction}
\label{sec:intro}

\begin{figure*}[tb!]
\begin{center}
\includegraphics[width=\linewidth]{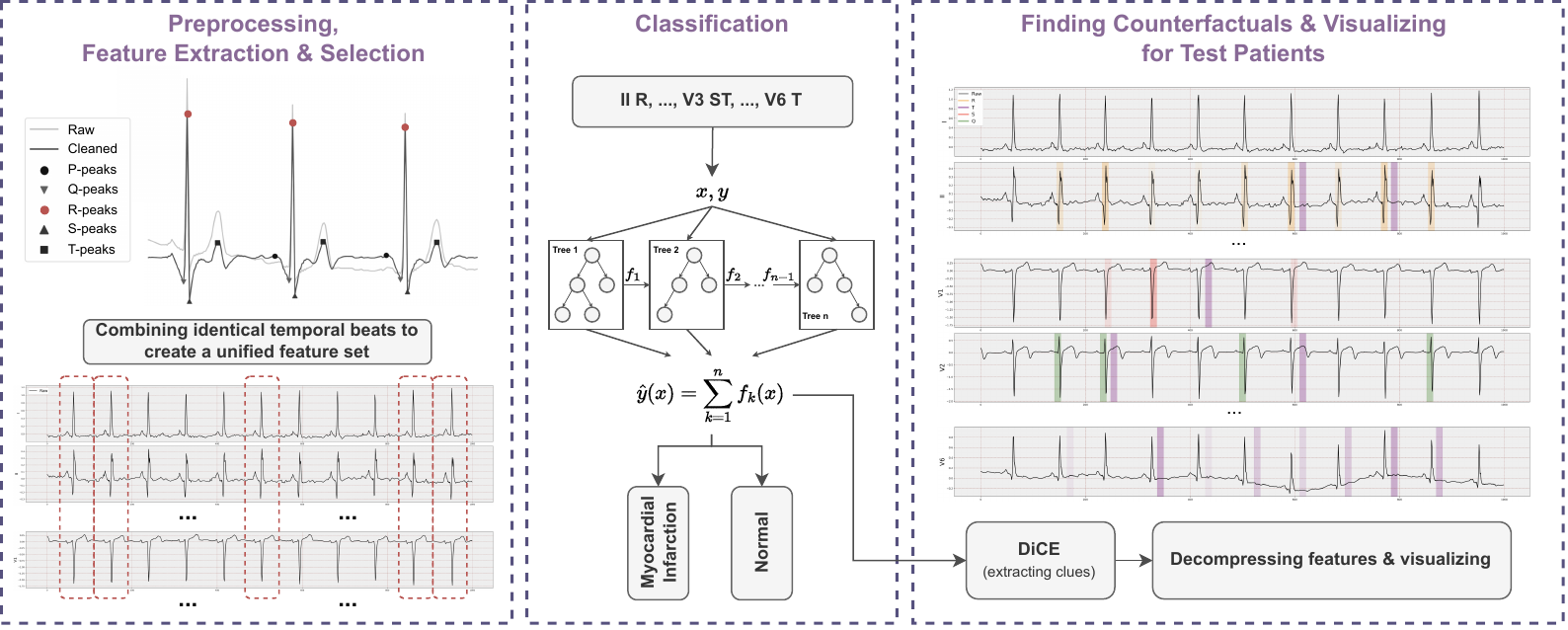}
\caption{Our proposed pipeline for interpretable ECG classification with improved visualization of counterfactuals has three steps: \textbf{1)} Signal preprocessing, extraction and compression of each cardiac cycle, was performed to create a feature set. \textbf{2)} The feature set, obtained through a data-driven and domain expertise, was employed to train the XGBoost model for classification. \textbf{3)} Counterfactual clues for individual patients were determined using the DiCE method, and the pertinent features were then decompressed and visualized on the ECG reports.} \label{Fig1}
\end{center}
\end{figure*}

Diagnosing myocardial infarction (MI) using electrocardiograms (ECG) remains challenging due to sometimes subtle changes in the ST segment, T wave, and QRS complex \cite{panju1998patient, thygesen2007universal}. When these changes are not evident, other tests are required for diagnosis. Early research on ECG-based diagnosis focused on machine learning (ML) with handcrafted features or Electronic Health Record (EHR) data \cite{dai2015prediction, sharma2018inferior, alizadehsani2019machine, ibtehaz2019vfpred, ayano2022interpretable}, utilizing explanation methods like LIME and SHAP to interpret model decisions \cite{ribeiro10.1145/2939672.2939778, lundberg2017unified, bodini2020interpretability, shetty2022machine}. \rtodo{R2.3}However, these methods may not fully meet clinicians' needs for practical, decision-supporting insights. Recent advancements have shifted towards deep learning, leveraging the entire 12-lead ECG signal for enhanced diagnostic accuracy \cite{hong2020opportunities, hasan2019deep, paragliola2021hybrid, gupta2021deep, wu2022deep}. Techniques like Grad-CAM have been employed to visually explain predictions \cite{balik2023interpretable, makimoto2020performance, jahmunah2022explainable, selvaraju2017grad}, though \rtodo{R2.3}challenges in accurately \textcolor{blue}{localizing} diagnostic signal regions remain.

In healthcare, cognitive biases and counterfactual thinking significantly influence medical decisions \cite{chen2023investigating, petrocelli2013pitfalls,  groopman2007doctors, durand2009causation}. Recent studies have explored statistical alternatives to instinct-based decisions \cite{gigerenzer2007helping, riffenburgh2020statistics}, proposing personalized treatments through counterfactual evaluations \cite{wang2021counterfactual, delaney2021instance, haldar2021reliable, xu2022counterfactual, nagesh2023explaining, todo2023counterfactual, tanyel2023beyond}.

\textcolor{blue}{The \rtodo{R1.1, R2.3}motivation for our study lies in addressing the limitations of existing interpretability techniques and improving practical usability in clinical settings. Existing explanation methods, such as LIME, SHAP, and Grad-CAM, though valuable, often fail to provide the depth and specificity required by clinicians to make informed decisions. In particular, these methods can struggle to clearly illustrate how specific changes in ECG signal characteristics influence diagnostic decisions, leaving a gap in their practical applicability. This gap underscores the need for more intuitive, actionable explanations that go beyond highlighting general areas of interest.}

Our study introduces a novel approach, utilizing synchronous heartbeat data across all ECG leads to develop a comprehensive feature set. This methodology, Visualizing Counterfactual Clues on Electrocardiograms (VCCE), aims to enhance disease detection and signal characteristic understanding across leads, providing personalized MI insights and counterfactual visualizations for clearer decision-making support (Figure \ref{Fig1}). \rtodo{R1.1}\textcolor{blue}{By focusing on both temporal and amplitude-based features, VCCE is designed to provide a holistic view of the cardiac signals, capturing subtle yet significant variations that are crucial for accurate MI diagnosis. Our approach also aims to bridge the gap between traditional black-box model predictions and the real-world needs of healthcare professionals by offering a tool that visually represents "what if" scenarios, thus providing deeper, patient-specific insights.}

\rtodo{R1.1}\textcolor{blue}{The proposed VCCE method integrates counterfactual reasoning to help clinicians understand how hypothetical modifications to specific ECG features could impact diagnostic outcomes. By offering an intuitive visualization of key features, our approach provides practical, easy-to-understand insights that align closely with the clinical workflow. This ultimately aims to improve both the accuracy of MI detection and the clinician's ability to confidently interpret and act on model outputs.}

\textcolor{blue}{Our contributions in this work are threefold:}
\begin{itemize}
    \item Streamlining feature extraction, ranking, and pruning, enriched with counterfactual insights for a deeper understanding of ECG signals.
    \item Offering a novel visualization of counterfactual interpretations for 12-lead ECG, facilitating practical application.
    \item Validating our methods and findings with evaluations from two cardiologists, demonstrating their clinical relevance.
\end{itemize}

\section{Previous Work on Counterfactuals}

\renewcommand{\arraystretch}{1.3}
\begin{table*}[t!]
\resizebox{\textwidth}{!}{
\begin{tabular}{cccccccccc}
\hline
\textbf{Paper} &
  \textbf{Year} &
  \textbf{Dataset} &
  \textbf{Sample} &
  \textbf{Task} &
  \textbf{Method} &
  \textbf{Level} &
  \textbf{\begin{tabular}[c]{@{}c@{}}Detailed ECG \\ Analysis\end{tabular}} &
  \textbf{\begin{tabular}[c]{@{}c@{}}Clinical \\ Validation\end{tabular}} &
  \textbf{\begin{tabular}[c]{@{}c@{}}Available \\ Code\end{tabular}} \\ \hline
Delaney et al. \cite{delaney2021instance} &
  2021 &
  ECG200 &
  200 &
  \begin{tabular}[c]{@{}c@{}}Time-series\\ Perturbation\end{tabular} &
  Native Guide &
  Beat &
  \xmark &
  \xmark &
  \cmark \\
Looveren et al. \cite{van2021conditional} &
  2021 &
  \cite{baim1986survival} &
  5000 &
  \begin{tabular}[c]{@{}c@{}}Time-series\\ Perturbation\end{tabular} &
  \begin{tabular}[c]{@{}c@{}}Conditional \\ Generative Models\end{tabular} &
  Beat &
  \xmark &
  \xmark &
  \xmark \\
Wang et al. \cite{wang2021learning}&
  2021 &
  TwoLeadECG &
  1162 &
  \begin{tabular}[c]{@{}c@{}}Time-series\\ Perturbation\end{tabular} &
  LatentCF++ &
  Beat &
  \xmark &
  \xmark &
  \cmark \\
Li et al. \cite{li2022motif} &
  2022 &
  ECG200 &
  200 &
  \begin{tabular}[c]{@{}c@{}}Time-series\\ Perturbation\end{tabular} &
  Motif-Guided &
  Beat &
  \xmark &
  \xmark &
  \xmark \\
Todo et al. \cite{todo2023counterfactual} &
  2023 &
  PTB-XL &
  21799 &
  \begin{tabular}[c]{@{}c@{}}Time-series\\ Perturbation\end{tabular} &
  \begin{tabular}[c]{@{}c@{}}Contrastive \\ VAE\end{tabular} &
  12-Lead &
  \xmark &
  \xmark &
  \xmark \\
Nagesh et al. \cite{nagesh2023explaining} &
  2023 &
  MIMIC-III &
  53423 &
  \begin{tabular}[c]{@{}c@{}}Feature\\ Perturbation\end{tabular} &
  \begin{tabular}[c]{@{}c@{}}Counterfactual \\ VAE\end{tabular} &
  EHR &
  \xmark &
  \xmark &
  \cmark \\
Ours &
  - &
  PTB-XL &
  3898 &
  \begin{tabular}[c]{@{}c@{}}Feature\\ Perturbation\end{tabular} &
  \begin{tabular}[c]{@{}c@{}} VCCE \\ (Visualization)\end{tabular} &
  12-Lead &
  \cmark &
  \cmark &
  \cmark
\end{tabular}}
\caption{Comparative overview of counterfactual analysis studies applied to ECG data. This table reviews various studies, all applying methods to ECG data, yet none deeply focus on ECG signals at a clinical level and lack clinical validation. The sample sizes vary, reflecting differences in dataset scales. Some datasets, like ECG200 and TwoLeadECG, are both small in size and scope, focusing on beat-level data. Looveren et al. extract only the first beat from each sample in Baim et al.'s dataset \cite{baim1986survival}. The PTB-XL dataset, utilized in both Todo et al.'s study and ours, is notably complex than other datasets. Time-series perturbation refers to alterations in the signal itself, whereas feature perturbation involves changes to features extracted from the signal or data sources like electronic health records (EHR). VAE (Variational Autoencoder).} \label{literature}
\end{table*}

Exploring human-centric explanations, counterfactuals consider hypothetical "what if" scenarios, offering insights across diverse domains, including philosophy, psychology, and AI \cite{miller2019explanation, starr2019counterfactuals}. This concept is particularly relevant in medicine, where understanding alternate outcomes could guide patient-specific decisions. AI research has progressively focused on counterfactual generation to personalize responses based on historical cases, thereby enhancing the explanatory depth of ML models \cite{wachter2017counterfactual, mothilal2020explaining, chou2022counterfactuals}. Guidotti's review \cite{guidotti2022counterfactual} and Verma et al.'s survey \cite{verma2020counterfactual} offer comprehensive overviews of counterfactual algorithms in ML, with growing interest in their application to ECG data \cite{delaney2021instance, li2022motif, todo2023counterfactual}. The DiCE algorithm, known for its user-friendliness, has been adopted for generating diverse counterfactual explanations \cite{mothilal2020explaining}.

Counterfactual reasoning extends to fields like finance and various other areas of medicine, where effective decision-making is crucial, highlighting the versatility of this approach in evaluating future possibilities for individuals \cite{fogel2006disposition, yang2020generating, hofler2005causal, wang2021counterfactual, waters2014grade, andini2017targeting}. Initial conceptual trade-offs involve the practicality of altering immutable characteristics, such as gender or age, for generating realistic scenarios. Nonetheless, recent algorithms have improved in creating plausible counterfactuals without resorting to unrealistic changes, also allowing the examination of potential biases related to gender, race, or age \cite{mikolajczyk2021towards}.

Table \ref{literature} summarizes key studies applying counterfactual methods to ECG data, providing a comparative perspective on datasets, methodologies, and validation approaches.

\section{Dataset}
Our study employed the PTB-XL dataset \cite{wagner2020ptb} (Version: 1.0.3) at a 100 Hz sampling rate. We balanced the dataset by randomly undersampling to match the number of normal (NORM) and MI cases, excluding patients with the ST-T changes (STTC) label. The training set comprised 1,559 NORM and 1,559 MI subjects, totaling 14,720 NORM beats and 16,356 MI beats. The test set included 390 subjects from each group, with 3,704 NORM beats and 4,134 MI beats. We excluded patients with subendocardial injury and posterior MI (PMI) due to diagnostic accuracy concerns and prioritized localized MI labels in cases of overlap, ensuring all included cases had 100\% certainty levels. The dataset was split 80\% for training and 20\% for testing, extracting 194 beat-level features across all 12 leads without considering age and gender to prevent bias. For clinical validation of visualized counterfactual clues, 51 random patients were analyzed.

\section{Methods}

\subsection{Preprocessing}

For the preprocessing of the raw ECG signals, we primarily utilized the NeuroKit2 library \cite{makowski2021neurokit2} (version 0.2.5) in Python (version 3.9.13). Both denoising and peak detection were accomplished using the method described in \cite{koka2022} (see Supplementary Figures C.1-C.4). \rtodo{R1.2}\textcolor{blue}{The denoising step was essential for removing noise artifacts that could interfere with subsequent analysis, while peak detection allowed for accurate identification of key waveform components, such as R-peaks, which are crucial for segmenting the ECG data into individual cardiac cycles.}

We excluded the initial and final beats to ensure the consistency of the utilized beat signals. Specifically, we removed these beats to ensure the extraction of RR\_prev and RR\_next features (Table \ref{T2}), which necessitate calculations before or after the R peak. While we could have assigned NaN or 0 values to these features, choosing to remove the corresponding beats is advantageous, because this approach mitigates opening and closing errors for the overall signal, hence also contributes to improved signal quality.

\begingroup
\color{blue}
\subsection{Feature Extraction}

\renewcommand{\arraystretch}{1.25}
\begin{table*}[htb!]
\centering
\resizebox{0.8\textwidth}{!}{
\begin{tabular}{@{}cllcl@{}}
\toprule
\multicolumn{2}{c}{\textbf{All Leads Common (Time Dependent)}} &
   &
  \multicolumn{2}{c}{\textbf{For Each Lead (Time Independent)}} \\ \midrule
\multicolumn{1}{r}{\textbf{\begin{tabular}[c]{@{}r@{}}RR\_Prev\\ RR\_Next\\ RR\_Rate\\ PR\_int\\ PR\_seg\\ QRS\\ P\_Wave\\ T\_Wave\\ T\_left\\ QT\\ QTc\\ ST\\ PT\\ PS\end{tabular}}} &
  \begin{tabular}[c]{@{}l@{}}Previous RR interval\\ Subsequent RR interval\\ RR\_Next / RR\_Prev\\ Time between P\_onset and R\_onset\\ Time between P\_offset and R\_onset\\ Time between R\_onset and R\_offset\\ Time between P\_onset and P\_offset\\ Time between T\_onset and T\_offset\\ Time between T\_onset and T\_peak\\ Time between R\_onset and T\_offset\\ Corrected QT according to Bazett’s formula\\ Time between R\_offset and T\_onset\\ Time between P\_onset and T\_offset\\ Time between P\_onset and S\_offset\end{tabular} &
   &
  \multicolumn{1}{r}{\textbf{\begin{tabular}[c]{@{}r@{}}R\\ P\\ Q\\ S\\ T\\ PQ\\ QR\\ RS\\ ST\\ PS\\ PT\\ QS\\ QT\\ ST\_mean\\ ST\_std\end{tabular}}} &
  \begin{tabular}[c]{@{}l@{}}R amplitude\\ P amplitude\\ Q amplitude\\ S amplitude\\ T amplitude\\ P\_a – Q\_a\\ Q\_a – R\_a\\ R\_a – S\_a\\ S\_a – T\_a\\ P\_a – S\_a\\ P\_a – T\_a\\ Q\_a – S\_a\\ Q\_a – T\_a\\ Mean(ST segment values)\\ STD(ST segment values)\end{tabular} \\ \bottomrule
\end{tabular}}
\caption{Extracted features. The left-hand side shows the temporal (time-dependent) features, while the right side shows the non-temporal (time-independent) features. The abbreviation 'Waveform\_a' denotes the amplitude of a waveform.} \label{T2}
\end{table*}

Once the ECG signals were preprocessed and segmented, we extracted features from each cardiac cycle. The feature extraction process was divided into two categories: temporal features and amplitude-based features.

\begin{description}
    \item[\textbf{Temporal Features}] 
    These features are consistent across all leads and are extracted based on the timing of key events within each cardiac cycle. Examples include:
    \begin{description}
        \item[RR\_Prev, RR\_Next:] The time intervals between successive R-peaks, indicating heart rate variability.
        \item[PR Interval:] The time between the onset of the P wave and the onset of the R peak.
        \item[QRS Duration:] The time between the onset and offset of the QRS complex.
        \item[QT Interval:] The time between the onset of the Q wave and the offset of the T wave.
    \end{description}
    These features were computed by measuring the time between characteristic points (e.g.,~P onset, R peak, T offset) identified during preprocessing. Temporal features provide insight into the overall rhythm and timing relationships within the heart's electrical activity.

    \item[\textbf{Amplitude Features}]
    These features are lead-specific and capture the voltage differences at characteristic waveform points, such as the P wave, QRS complex, and T wave. Examples include:
    \begin{description}
        \item[R Amplitude, P Amplitude, T Amplitude:] The peak voltage values of the R, P, and T waves, respectively, within each lead.
        \item[PQ, QR, RS:] Differences in amplitude between characteristic peaks, such as from P to Q, Q to R, and R to S.
        \item[ST\_mean, ST\_std:] The mean and standard deviation of the ST segment values.
    \end{description}
    These amplitude features were calculated by measuring the peak values and differences between peaks within each lead, allowing us to capture distinct characteristics of the electrical activity from different perspectives across the 12 leads. Amplitude features are particularly useful for detecting abnormalities in the morphology of the ECG signal, such as changes in the QRS complex or ST segment that may indicate myocardial infarction or ischemia.
\end{description}

By combining temporal features (which are uniform across leads) with amplitude features (which are distinct for each lead), we created a comprehensive feature set that captures both global timing information and lead-specific electrical activity. The complete set of extracted features, including descriptions, is presented in Table \ref{T2}.

This approach allowed us to develop a total of \rtodo{R2.7}194 features, representing a holistic view of the cardiac activity, including both the temporal sequence of events and the specific characteristics of each lead. This thorough feature extraction process is crucial for understanding the subtle differences in ECG signals, which can aid in diagnosing MI.
\endgroup

\begingroup
\color{blue}
\subsection{Feature Selection}

\rtodo{R1.3}To enhance ML model training and ensure the relevance of counterfactual scenarios by avoiding feature overload, we implemented a systematic feature selection process using recursive feature elimination (RFE) with the XGBoost (XGB) algorithm. This allowed us to rank features based on their importance and isolate the top 97 features that had the most significant impact on the model's performance. These top-ranked features were selected to prevent feature overload, which could otherwise negatively affect the model's efficiency and interpretability.

To further refine feature importance, these features were re-evaluated within the XGB framework, and the model was iteratively optimized by adding features one by one, starting from a smaller subset until the full set of 97 features was reached. This iterative evaluation helped us determine the optimal number of features needed for effective classification while maintaining model accuracy (See Supplementary Figure C.5).

For counterfactual generation, we selected feature sets in increments of 5, up to 20 features, since this process was conducted after the model development stage. Additionally, the final feature set was enhanced by blending domain knowledge with data-driven insights. Specifically, four key features were added to the top 20 to improve diagnostic precision for specific cases (Table
\ref{T3}). This combination of feature ranking, iterative evaluation, and domain expertise ensured that the final set of features was both relevant and effective for classification and counterfactual generation tasks.
\endgroup

\subsection{Counterfactual Generation Process}

\begin{figure*}[htb!]
\begin{center}
\includegraphics[width=\linewidth]{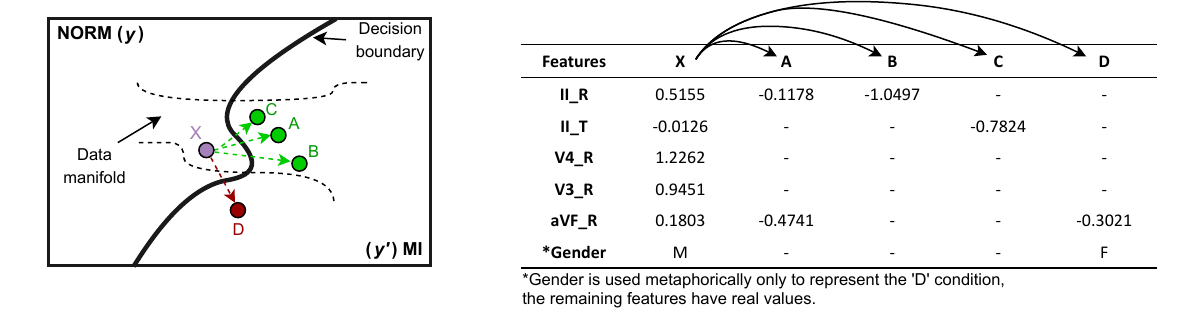}
\caption{The illustration of generating counterfactual explanations for NORM to MI case. The figure on the left depicts details of what happens during this process and the concept of "counterfactual
generation". On the right side, the illustration provides an example of the generated cases involved in the process (X to A, B, C or D). The green is for valid counterfactuals, while the red is representing fanciful generations. The symbol "-" signifies that the feature remained unchanged. Key ECG signal features are labeled: II\_R (R amplitude of lead II), II\_T (T amplitude of lead II), V4\_R (R amplitude of lead V4), V3\_R (R amplitude of lead V3), and aVF\_R (R amplitude of lead aVF). M: Male, F: Female. Data manifold represents real-life constraints, such as Gender.} \label{Fig2}\end{center}
\end{figure*}

Our counterfactual generation process (Figure \ref{Fig2}) integrates the DiCE ML library \cite{mothilal2020explaining} with a machine learning classifier, specifically an XGBoost model, to produce explainable AI outcomes. The process involves several key steps: classifier training, initialization of DiCE explainer, counterfactual generation, exploration, and data transformation for analysis.

\subsection{VCCE: Visualizing Counterfactual Clues on Electrocardiograms}

\begin{algorithm*}[t!]
\caption{VCCE Pseudocode}\label{algo1}
\begin{algorithmic}[1]
\Procedure{VCCE}{\text{signal}, \text{extracted\_features}, \text{counterfactuals}, \text{feature\_count}}
    \State \text{feature\_list} $\gets$ [\textit{P, Q, R, S, T, ST, PT, RS, QS}]
    \State \text{lead\_names} $\gets$ [\textit{I, II, III, aVR, aVL, aVF, V1, V2, V3, V4, V5, V6}]

    \State \text{combined\_data} $\gets$ \Call{PrepareData}{\text{counterfactuals}, \text{extracted\_features}}
    \State \text{visualization\_df} $\gets$ \Call{CreateVisualizationGraph}{\text{combined\_data}, \text{feature\_count}}

    \For{each \textit{lead} in \textit{lead\_names}}
        \For{each \textit{feature} in \textit{feature\_list}}
            \State \text{peaks} $\gets$ \Call{ExtractPeaks}{\text{visualization\_df}, \textit{lead}, \textit{feature}}
            \State \Call{PlotPeaks}{\text{peaks}, \text{signal}, \textit{lead}}
            \State \Call{PlotDottedLines}{\text{peaks}, \text{signal}, \textit{lead}} \Comment{For advanced features}
            \State \Call{PlotEmphasis}{\text{peaks}, \text{signal}, \textit{lead}}
        \EndFor
    \EndFor
    \State \Call{PlotECGReport}{\text{signal}, \text{visualization\_df}, $\ldots$} 
\EndProcedure
\end{algorithmic}
\end{algorithm*}

We have developed VCCE (Algorithm \ref{algo1}), a method to improve the presentation of counterfactual clues in ECG reports for clinicians. This method integrates counterfactual data with extracted ECG features, emphasizing significant waveform features (PQRST) and highlighting their diagnostic implications. VCCE employs a dynamic visualization approach, adapting to various counterfactual scenarios and their impacts on ECG waveforms.

For visualization, VCCE offers options to display peaks (such as the R peak) and chosen amplitude features (like II\_R), as identified by the counterfactual algorithm, in selected regions. Advanced features (for example, V3\_ST), which are calculated based on amplitude differences between two peaks, are represented as lines with their distance values overlaid on the ECG signal (refer to Supplementary Figure C.7). Additionally, in the visualization of counterfactuals, the frequency with which a particular amplitude feature is chosen in an alternative scenario directly influences its prominence in the ECG report. The more often a feature is selected, the more distinctly it is highlighted.

This approach, \rtodo{R2.17}adaptable beyond the specific ECG task, allows us to observe changes in the PQRST complex across multiple "possible worlds," offering a way to perceive alterations across various scenarios. The primary rationale for assuming the applicability of a VCCE-like approach across a range of ECG tasks lies in the nature of the ECG features employed, which are similar to those used in other ECG-dependent cardiac conditions. These features are extracted based on the structure of the P-, Q-, R-, S-, and T-waveforms (cardiac cycle) at the beat level, a consistent factor across all ECGs. The key advantage is that these derived features provide distinct temporal and amplitude-level information that is inherently interpretable. As a result, we can not only numerically analyze the generated counterfactual scenarios but also visually assess their impact on specific waves within particular leads and beats.

\rtodo{R2.6}In the proposed study, we note that the selected top 20 features are predominantly time independent due to their dependence on lead-wise amplitude characteristics across all MI cases in the dataset, with RR\_Next being the only temporal one. Therefore, our visualization method primarily focuses on amplitude-based (non-temporal) features. Additionally, we have included four more features (V2\_T, V3\_T, V4\_T, and V6\_T), selected by domain experts, to enhance individual resolution, such as distinguishing between inferior and anterior case differences. Consequently, we initiate a decoding process to unfold the temporally compressed beats. This decoding involves expanding the feature names to indicate the specific area influenced by the counterfactuals responsible for the change, encompassing details such as the lead, the waveform feature, and the beat associated with the counterfactual's impact. This way, we are individually marking the features that contribute the most significant distinction between the two diagnoses.

We utilize common metrics in ML (precision, recall, and F1 score), and present a set of custom evaluation metrics used to assess the plausibility of the generated counterfactual clues for ECG signals. The metrics are designed to capture both the clinical relevance of the counterfactual changes and the simplicity of the explanations. For detailed explanation of the metrics, see Supplementary Section A.

\section{Experimental Results}

We opted for the XGBoost model in this study because it maintains a balance across all metrics and offers the highest scores (refer to Supplementary Section B).

\subsection{Feature Elimination and XGBoost Classification}\label{fexgb}

For the training of the ML model, our initial step involved employing RFE on XGB to reduce the number of features. This process assigned a rank of 1 to 97 features, and we proceeded with our analysis using this subset of selected features. In subsequent stages of our analysis, we computed the importances of these chosen features within the XGB, as depicted in Supplementary Figure C.5a.

Following an in-depth examination of the selected features, we organized them based on their respective importance scores. Then, we conducted further training on XGB, progressively augmenting the number of features employed in the model (Supplementary Figure C.5b), where we can observe that having a mere 2\% score tolerance enables us to reduce the feature count from 86 to 19. This reduction brings benefits when working with counterfactuals.

\renewcommand{\arraystretch}{1.25}
\begin{table*}[htb!]
\begin{center}
\resizebox{0.9\textwidth}{!}{
\begin{tabular}{llllcc}
\cline{1-4} \cline{6-6}
\multicolumn{4}{c}{\textbf{Top 20 Features}}              &  & \textbf{4 Features Included by Clinicians} \\ \cline{1-4} \cline{6-6} 
\textbf{1)} II\_R: 24  & \textbf{6)} V1\_S: 15        & \textbf{11)} V1\_PS: 0   & \textbf{16)} V4\_PT: 2  &  & \textbf{21)} V2\_T: 24                                  \\
\textbf{2)} II\_T: 24  & \textbf{7)} aVR\_T: 5        & \textbf{12)} aVF\_QS: 0  & \textbf{17)} V2\_Q: 17   &  & \textbf{22)} V3\_T: 24                                  \\
\textbf{3)} V4\_R: 18  & \textbf{8)} V3\_ST: 22        & \textbf{13)} V1\_QS: 0   & \textbf{18)} V1\_T: 5   &  & \textbf{23)} V4\_T: 18                                  \\
\textbf{4)} V3\_R: 16 & \textbf{9)} aVF\_ST\_mean: 23 & \textbf{14)} V1\_ST: 19   & \textbf{19)} V5\_T: 20   &  & \textbf{24)} V6\_T: 19                                  \\
\textbf{5)} aVF\_R: 15 & \textbf{10)} V1\_RS: 6       & \textbf{15)} RR\_Next: 4 & \textbf{20)} aVL\_RS: 13 &  &  
\end{tabular}}
\end{center}
\caption{Top 20 features after our selection process, along with an additional 4 features chosen by clinicians. Moreover, we present the general clinical importance scores assigned by domain experts for each selected feature out of 24 points.\label{T3}}
\end{table*}

Through our experimental investigations, we discovered that utilizing 19 features is sufficient for extracting counterfactual instances. A lesser number of features is preferable for counterfactuals as it yields better and more reasonable explanations. To conduct a comprehensive assessment of counterfactuals, we generated distinct scenarios involving sets of 5, 10, 15, and 20 features. This evaluation encompassed considerations such as the time taken for generation (refer to Supplementary Figures C.6a and C.6b) and a comparative analysis of the most influential features affecting the outcomes. The F1 scores for each feature set are as follows: 81.40\% for 5 features, 83.50\% for 10 features, 85.83\% for 15 features, 86.59\% for 20 features, and 88.47\% for 97 features.
 
R-wave amplitude stands out as a prominent feature in the initial set of features (4 out of 5). We also observe the importance of T-wave amplitude in lead II and aVR, as well as S-wave amplitude in lead V1. Additionally, we notice T-waves with a lesser impact in V1 and V5 leads. In the case of V3 and V1, we observe various peak relationships, specifically the differences between S-amplitude and T-amplitude at V3, R-amplitude and S-amplitude, P-amplitude and S-amplitude, Q-amplitude and S-amplitude, as well as S-amplitude and T-amplitude at V1.

\subsection{Counterfactual Instance Generation} \label{cfe-gen}

To generate counterfactual instances, we proceeded with the chosen feature sets of 5, 10, 15, and 20 during the Section \ref{fexgb}. Table \ref{table2} provides a comprehensive overview of the outcomes, illustrating transformations from MI patients to NORM and vice versa. Specifically, a feature that is altered more frequently across different cases is considered more indicative or critical for changing the classification outcome. Thus, in the search for counterfactuals, those features that are most frequently chosen for alteration are highlighted as the most important or influential features in determining the class of a heartbeat. This insight is crucial for understanding which aspects of the data are most pivotal in the ML model's decision-making process and can guide efforts to improve model accuracy and interpretability.

In the counterfactual analysis conducted across a population of beats, it becomes evident that the XGB model more easily diagnoses beats from NORM patients compared to those from MI patients. This observation is supported by the pred/true ratio, which is detailed in Table \ref{table2}. Among the features under consideration, Feature 1-3 correspond to the top three features selected through the counterfactual optimization process. This selection indicates that these specific features exhibit better differentiation between NORM and MI cases, with distinct value distributions when we compare instances labeled as NORM to those categorized as MI. Consequently, we provided the most significant features by conducting additional feature selection across "possible worlds," for the collection of beats.

Analyzing the transition from NORM to MI, we observed notable changes in aVF\_R, V3\_ST, and II\_R among different feature sets. For the inverse transition, from MI to NORM, the features II\_T and II\_R underwent substantial changes.

\renewcommand{\arraystretch}{1.25}
\begin{table*}[tb!]
\centering
\resizebox{\textwidth}{!}{
\begin{tabular}{ccccclcccc}
\hline
                   & \multicolumn{4}{c}{\textbf{NORM to MI}}           &  & \multicolumn{4}{c}{\textbf{MI to NORM}}          \\ \cline{2-5} \cline{7-10} 
 &
  \textbf{5 features} &
  \textbf{10 features} &
  \textbf{15 features} &
  \textbf{20 features} &
  \textbf{} &
  \textbf{5 features} &
  \textbf{10 features} &
  \textbf{15 features} &
  \textbf{20 features} \\ \cline{2-5} \cline{7-10} 
\textbf{Pred/True} & 50/50      & 48/50      & 50/50      & 49/50      &  & 45/50      & 45/50     & 45/50      & 46/50      \\
\textbf{1st Significant} & aVF\_R: 60 & V3\_ST: 54 & II\_R: 37  & V3\_ST: 43 &  & II\_T: 66  & II\_T: 56 & II\_R: 60  & II\_R: 71  \\
\textbf{2nd Significant} & II\_R: 56  & V1\_S: 36  & V3\_ST: 37 & V2\_Q: 34  &  & aVF\_R: 43 & aVF\_R:50 & II\_T: 39  & V3\_R: 59  \\
\textbf{3rd Significant} & II\_T: 45  & II\_T: 35  & II\_T: 35  & II\_R: 31  &  & II\_R: 42  & II\_R:37  & aVF\_R: 38 & V4\_PT: 36 \\ \hline
\end{tabular}}
\caption{The table displays the key altered features across four distinct feature sets, accompanied by counterfactual cases from a sample of 50 heartbeats. In this context, an 'altered feature' is defined as a feature whose modification significantly influences the predicted class of a heartbeat in the ML model. The frequency at which a feature is altered in these counterfactual scenarios reflects its relative importance. 'Pred' represents the label assigned by the ML model, and 'True' indicates the actual class label of the heartbeats. This analysis focuses on heartbeats that the ML model accurately classified, excluding any counterfactuals from misdiagnosed cases. For each heartbeat, three counterfactual explanations are generated. For instance, if the feature set shows a pred/true ratio of 45/50, this leads to a total of 150 counterfactuals. However, considering 5 misdiagnoses, only 135 counterfactuals are used for statistical analysis. As an example, in the transition from MI to NORM, the II\_T feature changes 66 times within 135 possible scenarios for the selected 5 features set.} \label{table2}
\end{table*}

Generation times for counterfactual instances varied with the feature count. For NORM to MI transitions, generation times were 17.71 $\pm$ 33.0 seconds (5 features), 35.35 $\pm$ 45.62 seconds (10 features), 46.86 $\pm$ 53.35 seconds (15 features), and 58.06 $\pm$ 61.8 seconds (20 features). For MI to NORM transitions, generation times were notably higher: 73.2 $\pm$ 106.46 seconds (5 features), 132.64 $\pm$ 259.36 seconds (10 features), 111.94 $\pm$ 121.75 seconds (15 features), and 127.48 $\pm$ 129.02 seconds (20 features). This demonstrates that transitions from MI to NORM are more time-consuming and complex, reflecting the greater challenge and optimization effort required for these cases.

\subsection{Visualization of Counterfactual Clues}

We utilized I-1 to assess the quality of interpretability, I-2 to determine clinical alignment, I-3 to account for individual waveform errors corresponding to the validity score, and I-4 to evaluate clinical sparsity, which is expected to differ from XGBoost's sparsity based on feature importances. For detailed descriptions of the evaluation metrics I-1, I-2, I-3, and I-4, please refer to the Supplementary Section A.

In addition to the analysis in Section \ref{cfe-gen}, our preliminary empirical investigations on visual assessments have led domain experts to conclude that the incorporation of four additional features (V2\_T, V3\_T, V4\_T, and V6\_T) can be both beneficial and necessary. These features have not only enhanced the quality of visualization but also improved the classification accuracy. This was evident in the identification of local MI cases, such as Case ID 3234.0 in the dataset, where the system accurately identified 6 out of 7 individual beats, an improvement from the prior accuracy of 2 out of 7.

\begin{figure*}[tb!]
\begin{center}
\includegraphics[width=\linewidth]{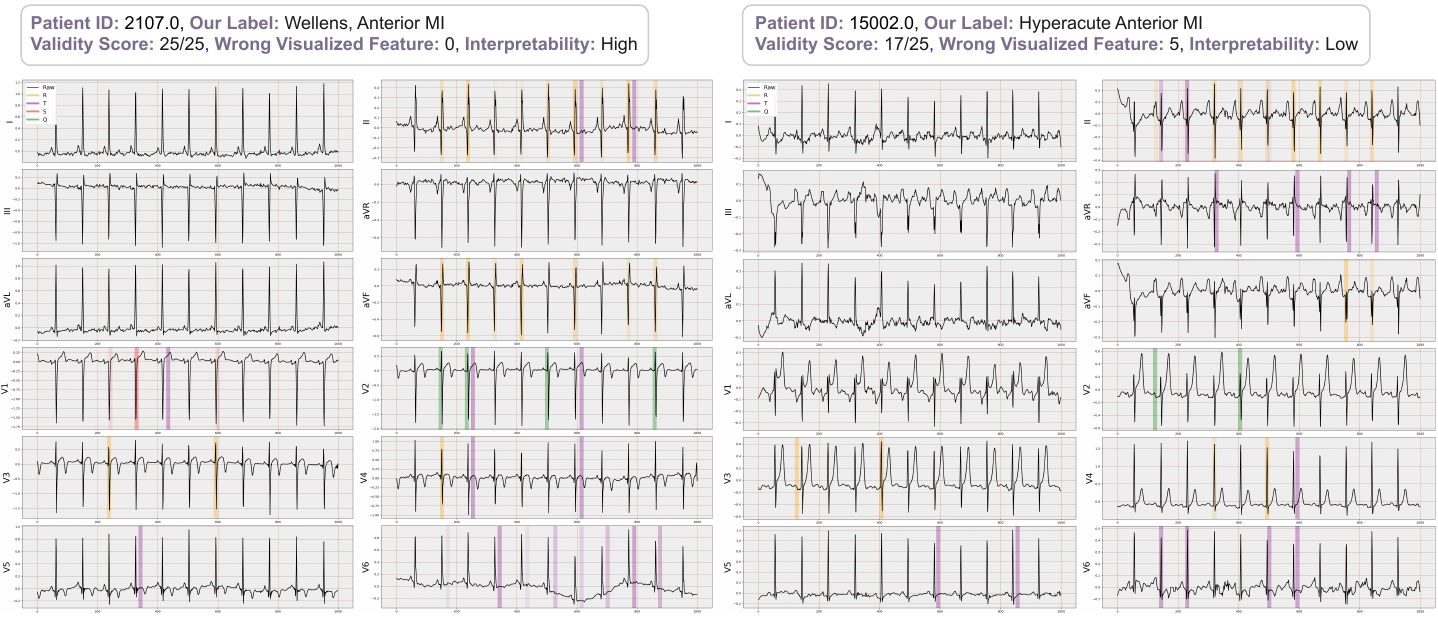}
\caption{Two example reports with visualized clues. Left: Patient's report with good and sufficient markings. Right: Patient's report with some unnecessary or incorrect markings, affecting interpretability scores.} \label{Fig4}
\end{center}
\end{figure*}

\begin{figure}[htb!]
\begin{center}
\includegraphics[width=\linewidth]{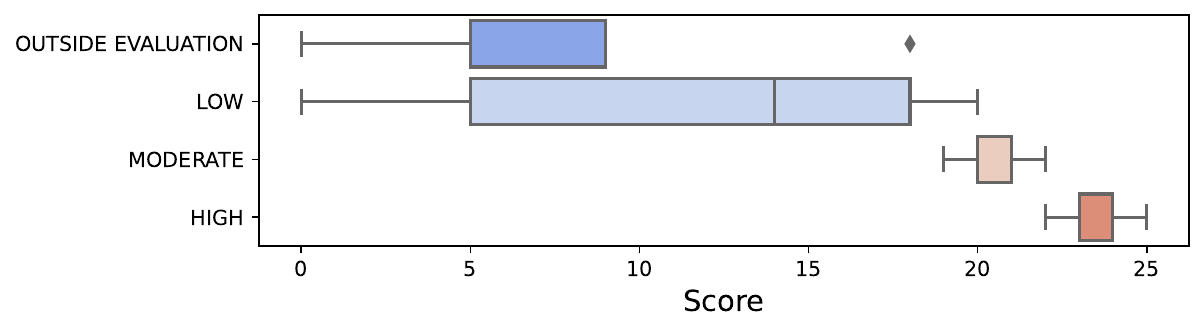}
\caption{Validation score distribution on interpretability.} \label{Fig5}
\end{center}
\end{figure}

51 reports were evaluated (Figure \ref{Fig5}); 5 of them were excluded due to extreme artifacts (n=1, $\mu$=0) and ventricular extrasystole (n=4, 10.25 ± 5.5). Of the remaining 46 reports, 17 were categorized as high (23.29 ± 1.04), 14 as moderate (20.28 ± 0.99), and 15 as low (11.20 ± 7.76) in terms of interpretability. The "±" symbol represents the mean and standard deviation of validation scores, respectively. A total of 31 reports were identified as acceptable.

\begin{figure*}[htb!]
\begin{center}
\includegraphics[width=\linewidth]{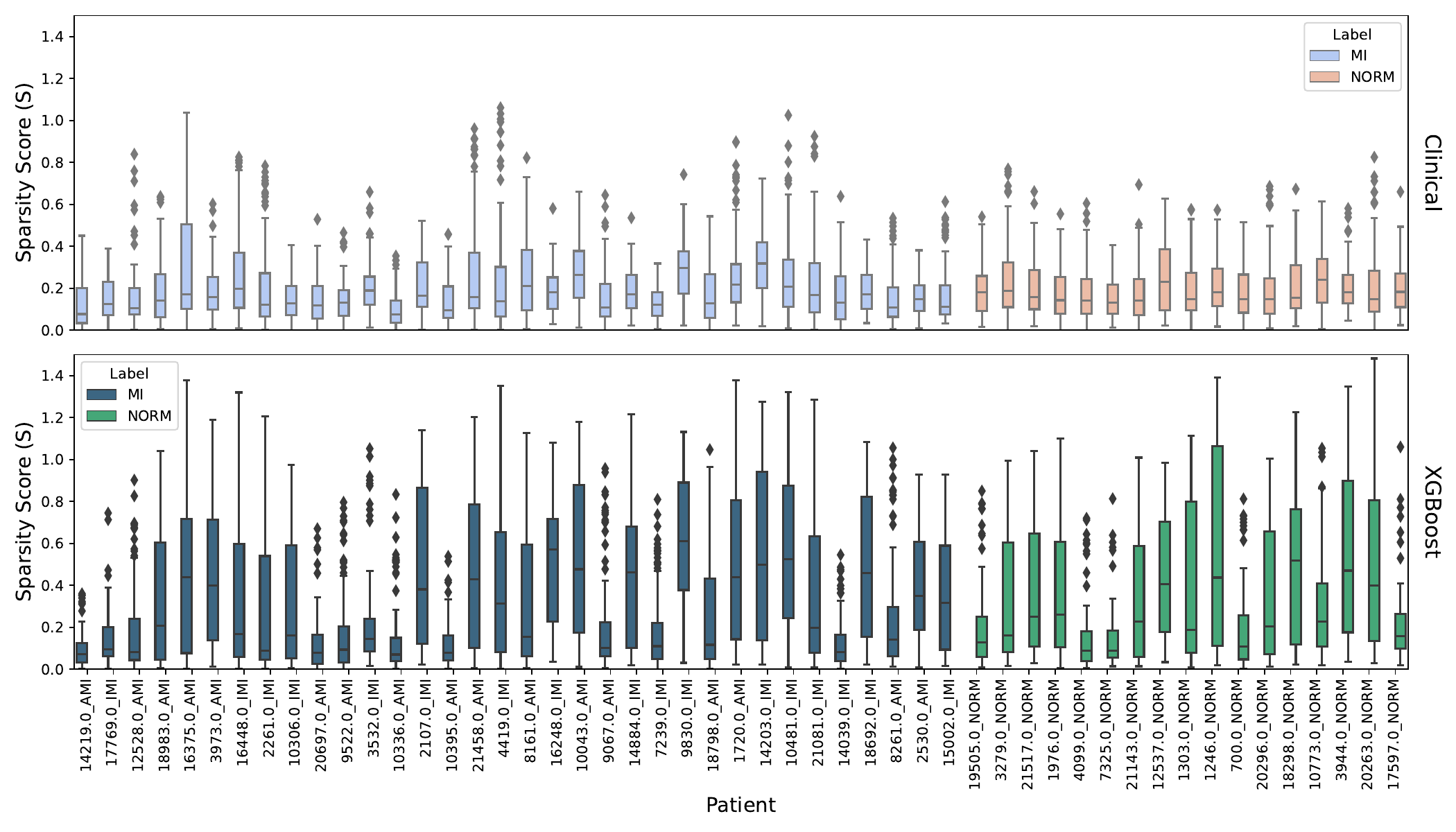}
\caption{Counterfactual sparsity scores. In this figure, patients with AMI and IMI are definitively diagnosed with Anterior MI and Inferior MI, respectively, according to the original dataset labels. NORM cases serve as the control group. Counterfactuals are weighted by clinical importance score and XGB feature importance, and we compared their sparsity scores. A higher sparsity score indicates a greater divergence from the original sample. This is because important features are selected more often, resulting in a more diverse counterfactual sample. The box plot represents all beats for patients and the corresponding counterfactual for each beat.} \label{Fig6}
\end{center}
\end{figure*}

In terms of overall counterfactual sparsity, by using our clinical scoring (I-4), counterfactuals demonstrated lower sparsity (0.19) compared to the XGB feature importances as scoring, which had a sparsity of 0.33. Patient-wise differences can be observed by comparing clinical and XGB boxplots. The general sparsity in XGB is evident in the Figure \ref{Fig6}, although some cases displayed even lower sparsity than their clinical-weighted counterparts.

In the alignment of blind evaluations and identified clues, we obtained scores of 0.83 ± 0.12 for high-quality reports, 0.57 ± 0.10 for moderate-quality reports, and 0.22 ± 0.24 for low-quality reports. We observed that a many of the low-quality reports were attributed to incorrect peak detection, which inevitably resulted in their low-quality classification. Moreover, these low-quality reports often contained labels that were inconsistent with our blind evaluations and the original labels. 

For more detailed clinical explanations of some of the selected cases, please refer to Figures C.8, C.9, and C.10 in the Supplementary Material.

\section{Discussion}

AI integration into healthcare faces challenges, notably clinician concerns over AI decision-making. Despite advancements, making AI decisions understandable is difficult. Our study focuses on explainable AI (XAI), specifically counterfactual explanations, to demystify ML decisions in ECG analysis with VCCE.

We chose the XGB algorithm for its superior performance (89.36\% AUC) on the PTB-XL dataset after comparing several ML algorithms. The XGB model's max\_depth parameter helps avoid overfitting, and we optimized to maintain performance with minimal complexity, selecting 20 features while considering dataset noise and ensuring high model accuracy.

To \rtodo{R2.5}address potential bias driven by data, we incorporated four features recommended by clinicians to enhance local clues. Counterfactual explanations refine feature selection by adding minimal yet impactful modifications, which improve model performance and provide a more nuanced understanding of the data.

Our model, with 86.3\% accuracy using 19 features, faces challenges with ventricular extrasystole (VES) and bundle branch blocks due to their unique morphologies. The reliance on standard evaluation criteria for these conditions can lead to inaccuracies. The PTB-XL dataset's label reliability is also questioned. While most studies accept the labels as they are \cite{smigiel2021ecg, wang2021automated}, our blind evaluations identified numerous undecidable cases from ECG readings, and there were potentially wrongly interpreted patients.

\textcolor{blue}{Blind \rtodo{R2.5}evaluations of counterfactual explanations showed they provide meaningful insights without contradictions, supported by expert annotations. Our method's use of sparsity and cognitive interpretability scores (17 high, 14 moderate, 15 low) alongside validity scores (0.83 for high, 0.57 for moderate, and 0.22 for low) demonstrates the effectiveness and promise of counterfactual explanations in enhancing understanding of ML decisions in ECG analysis.}

\textcolor{blue}{In counterfactual research,\rtodo{R2.5} including ECG analysis, diverse algorithms have been developed, focusing on signal perturbation within short, restricted beats \cite{delaney2021instance, wang2021learning, li2022motif}. Our approach uniquely analyzes entire 10-second beats from 12-lead ECGs collectively, offering comprehensive insights into the entire ECG report. Despite generative methods' potential, existing research \cite{van2021conditional,todo2023counterfactual} often lacks clinical grounding and open-source code, with some studies \cite{nagesh2023explaining} not prioritizing ECG analysis. This reflects a broader issue: current studies fail to align with clinical needs, using metrics like sparsity and validity that hold limited medical relevance due to the lack of suitable datasets for their application. Addressing this gap, our work introduces new metrics designed for medical contexts and demonstrates the clinical importance of analyzing 12-lead counterfactuals. However, communicating these findings to clinicians effectively remains challenging with existing methods. Our innovative approach, validated by cardiologists, presents a promising direction for future research, illustrating the potential to bridge the gap between counterfactual research and practical clinical application.}

\textcolor{blue}{This\rtodo{R2.5} study has several limitations that offer directions for future research. The accuracy of results may be impacted by the variability in peak detection performance on amplitude values, addressed partially by employing the \cite{koka2022} method. However, high-noise and irregular heart rhythm scenarios remain challenging. The absence of visible active heart rhythm waves in some beats due to external factors during ECG recordings could lead to inaccurately indexed beats; this was mitigated by excluding such beats during visualization. Additionally, the study did not assess the entire signal in one go but analyzed all leads for each heartbeat to determine patient outcomes. Lastly, the optimization duration of the DiCE method during counterfactual generation poses a limitation, but exploring alternative algorithms may offer solutions. These limitations provide a foundation for enhancing visualization methods and exploring diverse features in future works.}

\section{Conclusion}

Our study proposes a novel idea for improving ECG analysis by integrating counterfactual explanations with our VCCE visualization technique. By merging AI insights with expert cardiological evaluations, we offer a perspective on MI analysis. Our approach, validated through custom metrics and comprehensive evaluations, shows potential for enhanced visualization and diagnostic precision. This research provides a stepping stone for future work aimed at refining AI applications and addressing cognitive needs in medical contexts.

\section*{Funding}
This study was supported by the Health Institutes of Turkiye (TUSEB) 2022-EKG-01 Program (Project No: 20101).



\bibliographystyle{elsarticle-num-names} 
\bibliography{ref}






\section*{Supplementary File}

\appendix

\counterwithin{table}{section}
\renewcommand\thetable{\thesection.\arabic{table}}

\counterwithin{figure}{section}
\renewcommand\thefigure{\thesection.\arabic{figure}}

\section{Evaluation Metrics}\label{eval}

We utilize common metrics in ML (precision (Eq. \ref{eq:precision}), recall (Eq. \ref{eq:recall}), and F1 score (Eq. \ref{eq:f1score})), and present a set of custom evaluation metrics used to assess the plausibility of the generated counterfactual clues for ECG signals. The metrics are designed to capture both the clinical relevance of the counterfactual changes and the simplicity of the explanations.

In the custom evaluation, we initiated a blind evaluation by providing domain experts with raw ECG reports to establish a baseline and enrich our visual evaluation with additional data. Subsequently, we issued four primary instructions to the domain experts for validating the generated visual reports. In the first evaluation, we requested them to assess the report's interpretability and quality as a cognitive task. In the second evaluation, we asked them to identify individual errors in selecting clues for the visualization and to evaluate the alignment with a blind evaluation. For the third evaluation, we requested them to rate the clinical relevance of our visualized clues on a scale from 0 to 5, which we considered as a visualization validity score. Finally, we asked them to assess the overall clinical significance of the selected features to determine the weighted sparsity of the counterfactuals. We have also made the labels assigned by clinicians for specific instructions available on our GitHub page \href{https://github.com/tanyelai/vcce}{https://github.com/tanyelai/vcce}.

\begin{figure}[htb!]
\begin{center}
\includegraphics[width=0.85\linewidth]{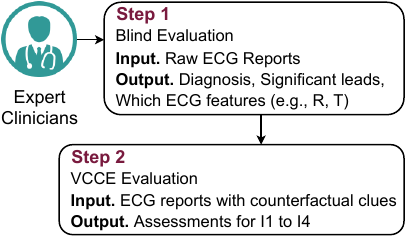}
\caption{Evaluation process of the domain experts.}\label{FigEval}\end{center}
\end{figure}

\textbf{\textit{Precision, Recall and F1 Score}}. We assessed the ML models using Precision, a measure of correct positive predictions; Recall, the model's ability to identify all positives; and the F1 Score, a balance between Precision and Recall, especially useful in imbalanced datasets. In classification tasks, True Positives (\text{TP}) are instances correctly predicted as positive, False Positives (\text{FP}) are instances incorrectly predicted as positive, and False Negatives (\text{FN}) are instances incorrectly predicted as negative. The validation metrics used in ML are as following equations:
\begin{center}
\begin{minipage}[t!]{0.4\textwidth}
    \centering
    \begin{gather}
        \text{Precision} = \frac{\text{TP}}{\text{TP} + \text{FP}} \label{eq:precision}
    \end{gather}
\end{minipage}%
\end{center}
\begin{center}
\begin{minipage}[t!]{.4\textwidth}
    \centering
    \begin{gather}
        \text{Recall} = \frac{\text{TP}}{\text{TP} + \text{FN}} \label{eq:recall}
    \end{gather}
\end{minipage}%
\end{center}
\begin{center}
\begin{minipage}[t!]{0.4\textwidth}
    \centering
    \begin{gather}
        \text{F1 Score} = 2 \times \frac{\text{Precision} \times \text{Recall}}{\text{Precision} + \text{Recall}} \label{eq:f1score}
    \end{gather}
\end{minipage}
\end{center}

\textbf{\textit{Directions to Measure Custom Metrics}}. The simplified instructions for cardiologists consist of two main parts: blind evaluation and visualized clues evaluation (Figure \ref{FigEval}). Each patient has two ECG reports: an original and a marked one. The original reports will be used to assess clinical consistency, while the marked reports will be used to evaluate cognitive aspects of the method, including its `understandability,' `interpretability,' and `plausibility.'
\begin{enumerate}
    \item \textbf{Blind:} A diagnosis will be established based on the original ECG; this diagnosis will outline which leads are significant and which cardiac characteristics are being assessed clinically.
    
    \item \textbf{Visualized Clues:} We test the clarity and accuracy of the marked areas with our method under the following four instructions (I):
    
    \begin{enumerate}
        \item[\textbf{I-1:}] Can the meanings of the markings be understood? Evaluate their interpretability and quality as `low,' `moderate,' or `good'.
        
        \item[\textbf{I-2:}] Does it make sense for the mark to be here? Please also provide the total count of incorrect marks in the given report.

        \item[\textbf{I-3:}] Are the marked signal features clinically significant (if not marked, write "None")? If they are significant, please rate their importance between 1-5. If they are not significant but are marked, enter 0. If they are not significant and are also not marked (indicating the correct decision of non-selection), reward this accuracy by entering 5.
        
        \item[\textbf{I-4:}] In order to evaluate overall feature selection, please rate how clinically significant our chosen 24 features are for you on a scale of 0-24.
    \end{enumerate}  
\end{enumerate}

\textbf{\textit{Clinical Interpretability Score (I-1)}}. reflects the ease with which the clinicians can understand and interpret the counterfactual clues. This score is a qualitative assessment provided by the clinicians, indicating the level of interpretability (i.e., `low,' `moderate,' `good').

\textbf{\textit{Clinical Outcome Alignment Score (I-2)}}. quantifies the extent to which the predicted clinical outcomes of the counterfactual explanations align with the experts' assessments of potential clinical outcomes. Given the cardiologists' list of important leads obtained through blind evaluation, we introduce a weight \(W_i\) for each lead, where \(W_i = 3\) if the \(i^{th}\) lead is in the important lead list, and \(W_i = 1\) otherwise. The alignment (A) score is then calculated using counterfactuals (cf) as follows:
\begin{align}
\text{A} = \frac{\sum_{i=1}^{n} (W_{i} \cdot \text{Indicator}(\text{Lead } i \text{ aligned with cf}))}{\max(\sum_{i=1}^{n} W_{i}, 1)}
\end{align}
Where:
\begin{itemize}
    \item \(n\) is the total number of leads considered.
    \item \(W_{i}\) is the weight of the \(i^{th}\) lead, which is 3 if the lead is in the cardiologists' important lead list, and 1 otherwise.
    \item The Indicator function returns 1 if the \(i^{th}\) lead is aligned with the counterfactuals, and 0 otherwise.
\end{itemize}

The denominator \(\max(\sum_{i=1}^{n} W_{i}, 1)\) ensures that the alignment score is normalized to the interval \([0, 1]\), even in cases where the weight of important leads could potentially make the score greater than 1.

\textbf{\textit{Visualization Validity Score (I-3)}}. gauges the clinical relevance of generated visual clues displayed on the ECG. The Visualization Validity Score (VVS) for each patient is the summation of scores from individual visual clues, offering insight into the validity of the VCCE.

Given visual clues based on the P, Q, R, S, T waveforms, the VVS for each patient is formulated as:
\begin{equation}
\text{VVS} = \sum_{i=1}^{5} \text{Score}_{i}
\end{equation}
where \(\text{Score}_{i}\) represents the score assigned to the \(i^{th}\) waveform by the clinician.

\textbf{\textit{Feature Importance Weighted Sparsity Score (I-4)}}. This metric is designed to measure both the influence and sparsity of each ECG feature when generating counterfactual explanations. By evaluating the interaction between the feature's sparsity and its clinical or XGBoost-derived importance, it allows for more insightful and reliable interpretations.

Given $m$ counterfactuals, $n$ ECG features, and the respective weights $w_{mi}$ (clinical or XGBoost) for the $i$th feature, the sparsity scores $S_{\text{method}, j}$ for the $j$th counterfactual are computed using deviations $x_{ij}$ from the original instance. The overall mean and standard deviation for each sparsity score method are denoted by $\overline{S}_{\text{method}}$ and $\sigma_{S_{\text{method}}}$ respectively.

The sparsity scores based on the method can be represented as:
\begin{align}
    \overline{S}_{\text{method}} &= \frac{1}{m} \sum_{j=1}^{m} \left( \sum_{i=1}^{n} |x_{ij} \cdot w_{mi}| \right), \\
    \sigma_{S_{\text{method}}} &= \sqrt{\frac{1}{m} \sum_{j=1}^{m} \left( \sum_{i=1}^{n} |x_{ij} \cdot w_{mi}| - \overline{S}_{\text{method}} \right)^2}
\end{align}
where method $\in$ \{clinical, XGBoost\} and $w_{mi}$ is the weight corresponding to the chosen method.

A higher value of $\overline{S}$ indicates the importance of specific ECG features in counterfactual explanations, while the associated $\sigma$ quantifies variability across different instances.

\section{Model Selection}

Table \ref{AT1} illustrates the comparative performance of several machine learning models applied to our dataset. Among the evaluated models, XGBoost emerged as the most proficient, achieving the highest scores across all metrics, including Accuracy (89.44\%), Precision (89.74\%), Recall (87.99\%), F1-Score (88.73\%), and AUC-ROC (89.36\%). In contrast, the Gaussian Naive Bayes model exhibited the lowest accuracy and precision, yet it achieved a relatively high recall of 84.77\%. The ensemble models such as Random Forest, Gradient Boosting, AdaBoost, and Extra Trees, along with the Support Vector Machine, demonstrated decent efficacy. Furthermore, Logistic Regression and K-Nearest Neighbors provided moderate reliability and scores. We opted for the XGBoost model in this study because it maintains a balance across all metrics and offers the highest scores.

\renewcommand{\arraystretch}{1.25}
\begin{table*}[tb!]
\centering
\resizebox{0.92\textwidth}{!}{
\begin{tabular}{cccccc}
\textbf{Model Name} & \textbf{Accuracy} & \textbf{Precision} & \textbf{Recall} & \textbf{F1-Score} & \textbf{AUC-ROC} \\ \hline
XGBoost$^\ast$             & \textbf{89.44}    & \textbf{89.74}     & \textbf{87.99}  & \textbf{88.73}    & \textbf{89.36}   \\
AdaBoost               & 84.80        & 84.74        & 82.75        & 83.73        & 84.70        \\
Gradient Boosting      & 87.74        & 88.49        & 85.12        & 86.78        & 87.60        \\
Extra Trees            & 87.67 (0.08) & 88.28 (0.22) & 85.23 (0.13) & 86.73 (0.06) & 87.55 (0.07) \\
Random Forest          & 87.68 (0.07) & 89.03 (0.13) & 84.31 (0.27) & 86.61 (0.10) & 87.50 (0.08) \\
Logistic Regression    & 83.85        & 83.27        & 82.37        & 82.82        & 83.77        \\
Support Vector Machine & 87.96        & 88.55        & 85.58        & 87.04        & 87.83        \\
Gaussian Naive Bayes   & 74.57        & 68.72        & 84.77        & 75.91        & 75.10        \\
K-Nearest Neighbors    & 82.57        & 79.78        & 84.56        & 82.10        & 82.68       
\end{tabular}}
\caption{Machine learning models' outcomes for the utilized dataset on different trials, with the standard deviation of randomly initialized trees shown in brackets. XGBoost$^\ast$ results denote statistically significant differences compared to other models (p-value < 0.01). We applied McNemar's test, which is designed to assess whether there is a significant difference in the proportions of correct and incorrect predictions between the two models.}\label{AT1}
\end{table*}
\newpage

\section{Figures}
\begin{figure*}[htbp]
\centering
\includegraphics[width=\linewidth]{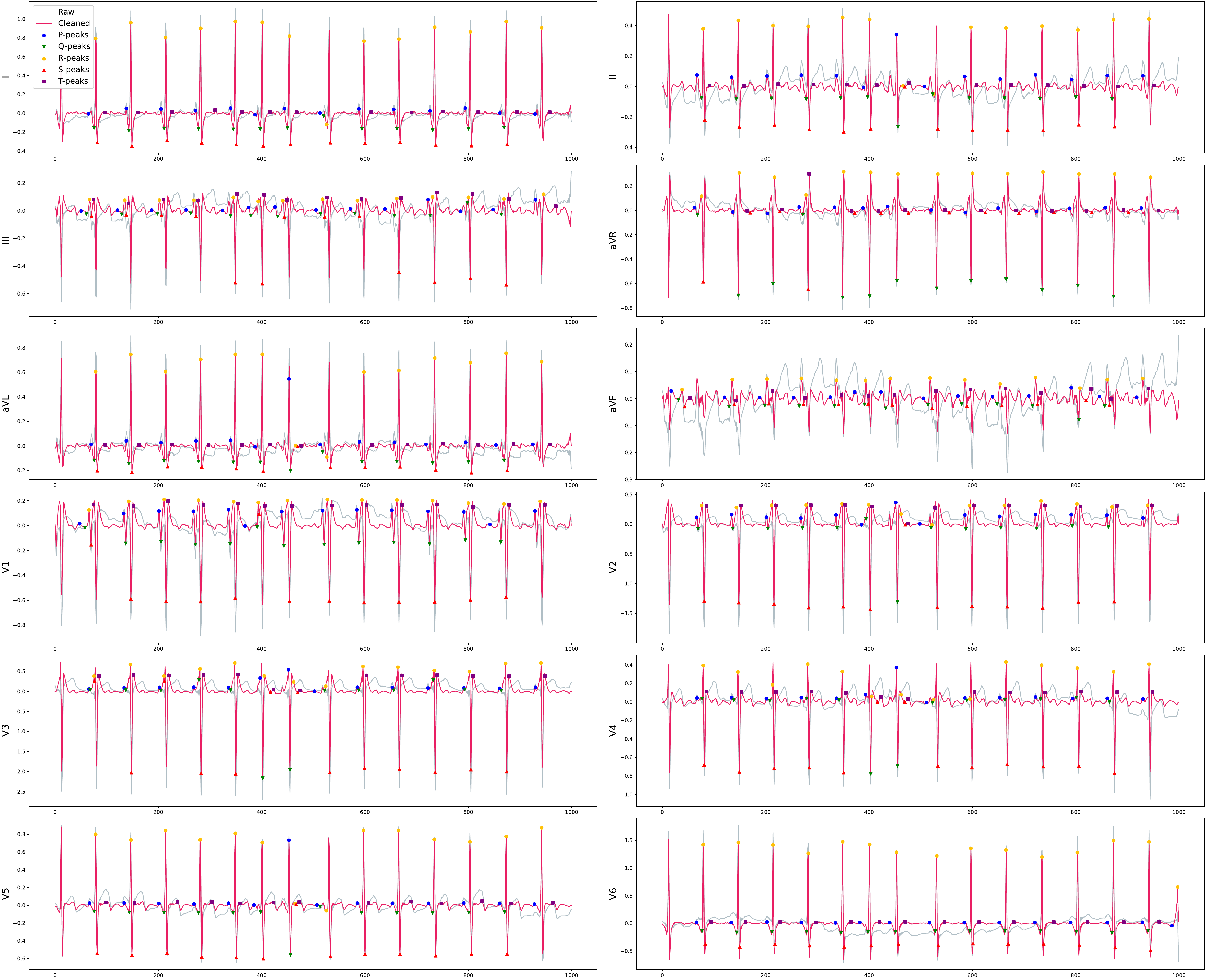}
\caption{Denoising and peaks detection using \texttt{koka2022} method for a random MI patient.}
\label{Afig:1}
\end{figure*}

We do not present numerical results for denoising and peak detection as they are out of scope; however, we conducted experimental assessments during method selection. We explored various techniques, and their efficacy varied across different random patients—some were beneficial while others were not. Certainly, we found no universally applicable method that effectively identified peaks in all the patients we studied. In patients with a substantial amount of noise, the selection of peaks can deviate, impacting the outcomes, and this deviation is inevitable for most patients, even with methods that are substantiated in the current literature.

\begin{figure*}[htbp]
\centering
\includegraphics[width=\linewidth]{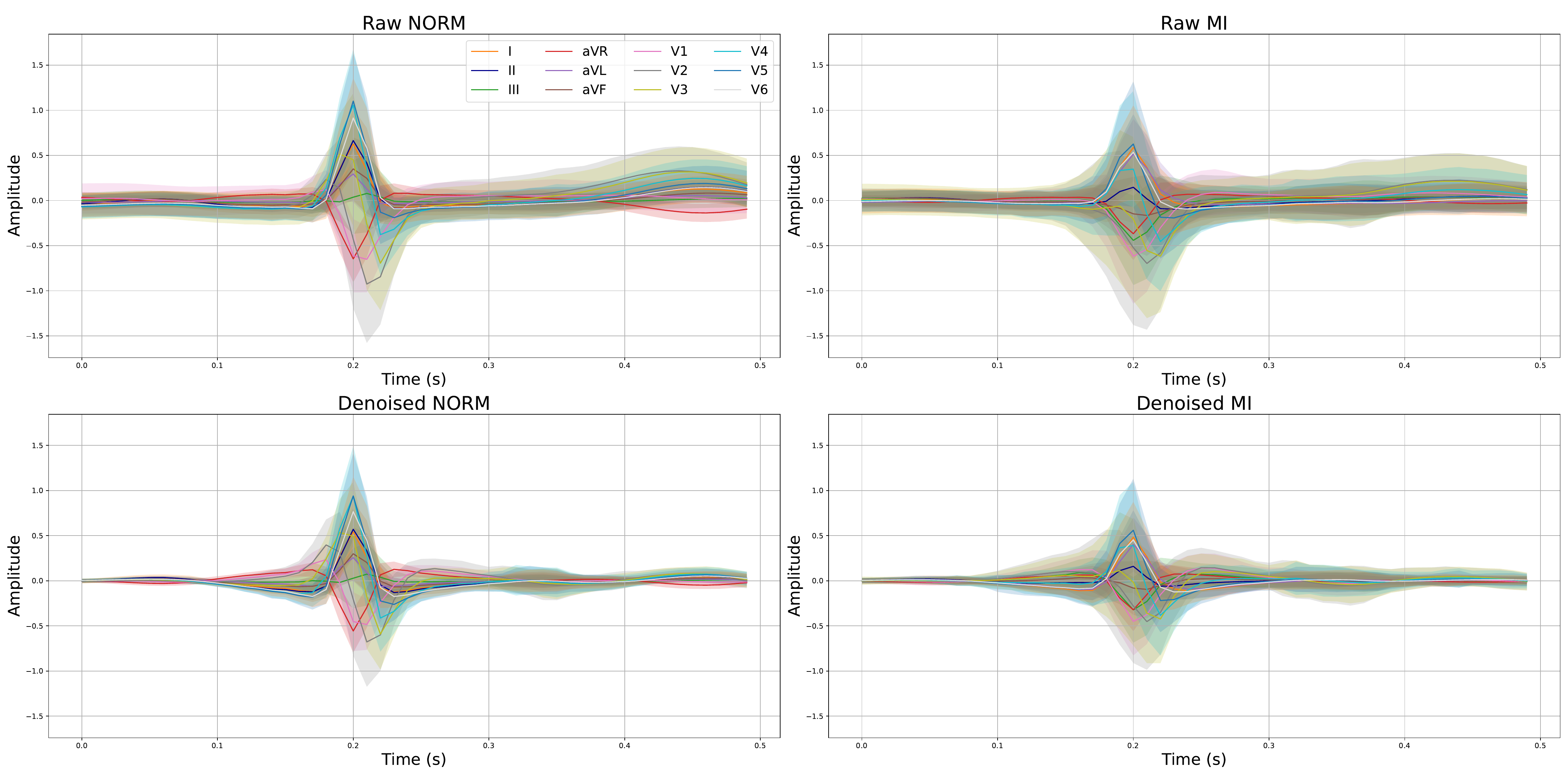}
\caption{The mean and standard deviation of 12-lead data are compared before and after preprocessing using the ``koka2022" method at one beat. The lines serve as references for the population mean, while the emphases are on illustrating the standard deviation.}
\label{Afig:2}
\end{figure*}

The Figure \ref{Afig:2} illustrates the distribution of raw and denoised signals across various leads for a selected beat within the population. It is important to note that individual investigations might encounter significant noise, and this figure serves as an example to showcase the substantial variance in beats across the population. Moreover, the figure visibly demonstrates the considerable discrepancy in signal amplitude values between the NORM and MI populations. 

\begin{figure*}[htbp]
\centering
\includegraphics[width=\linewidth]{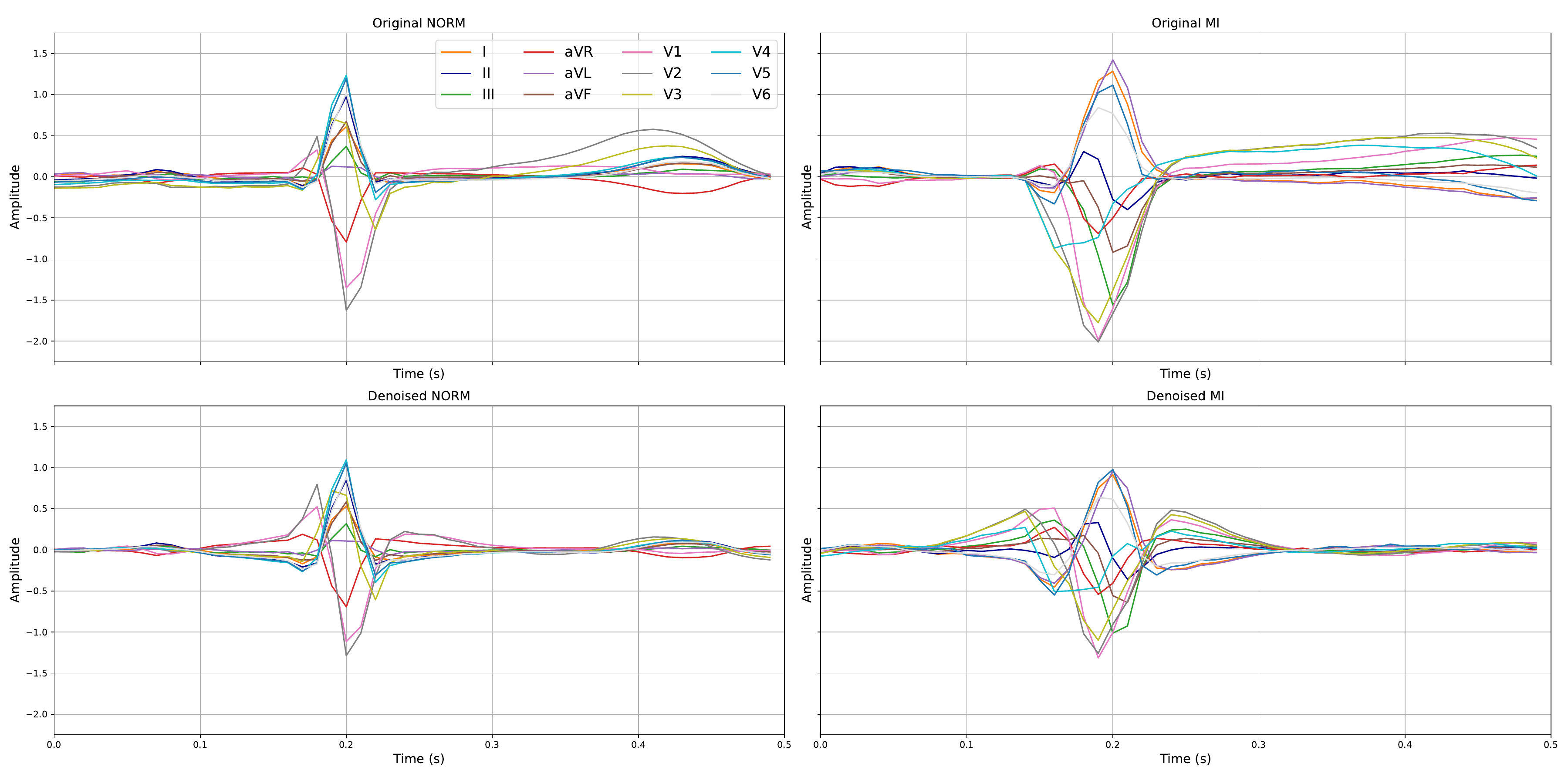}
\caption{Impact of denoising on one random NORM and MI patients on third beat.}
\label{Afig:3}
\end{figure*}

\begin{figure*}[htbp]
\centering
\fbox{\includegraphics[width=0.82\linewidth]{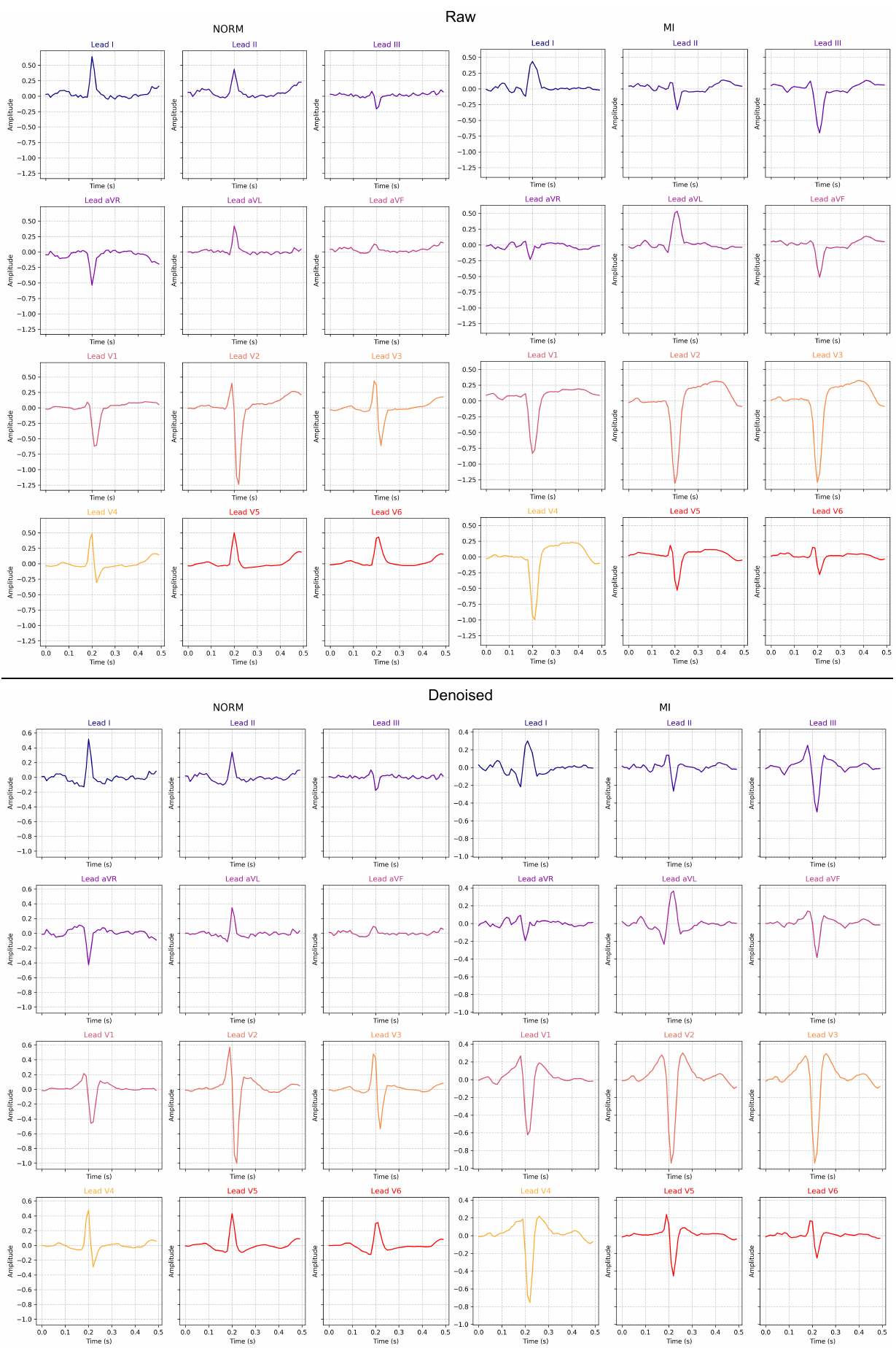}}
\caption{Beat-level example before and after denoising for one random NORM and MI patients.}
\label{Afig:4}
\end{figure*}

The effect of denoising on one beat is demonstrated for two random samples in Figure \ref{Afig:3}. Moreover, in Figure \ref{Afig:4}, we can also observe how the signal in each lead changes discretely for the samples.

\begin{figure*}[htbp]
    \centering
    \begin{subfigure}[b]{\linewidth}
        \centering
        \includegraphics[width=\linewidth]{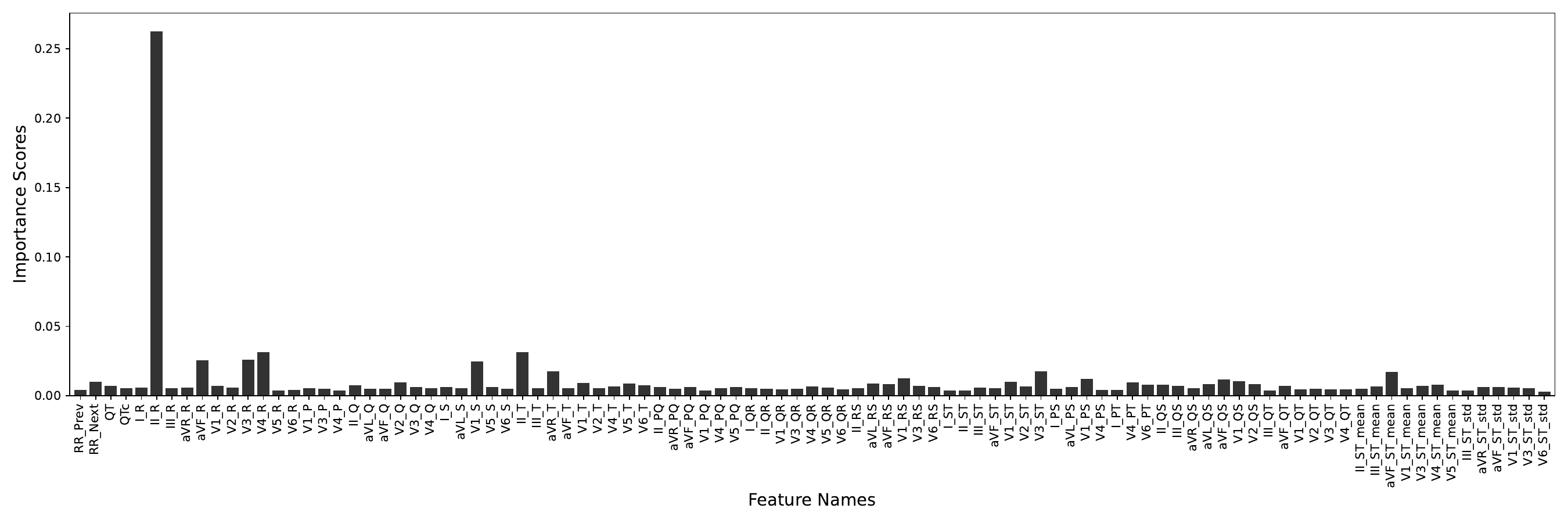}
        \caption{The importance report for each chosen feature in the classification of NORM and MI.}
        \label{Afig:5}
    \end{subfigure}
    \hfill
    \begin{subfigure}[b]{\linewidth}
        \centering
        \includegraphics[width=\linewidth]{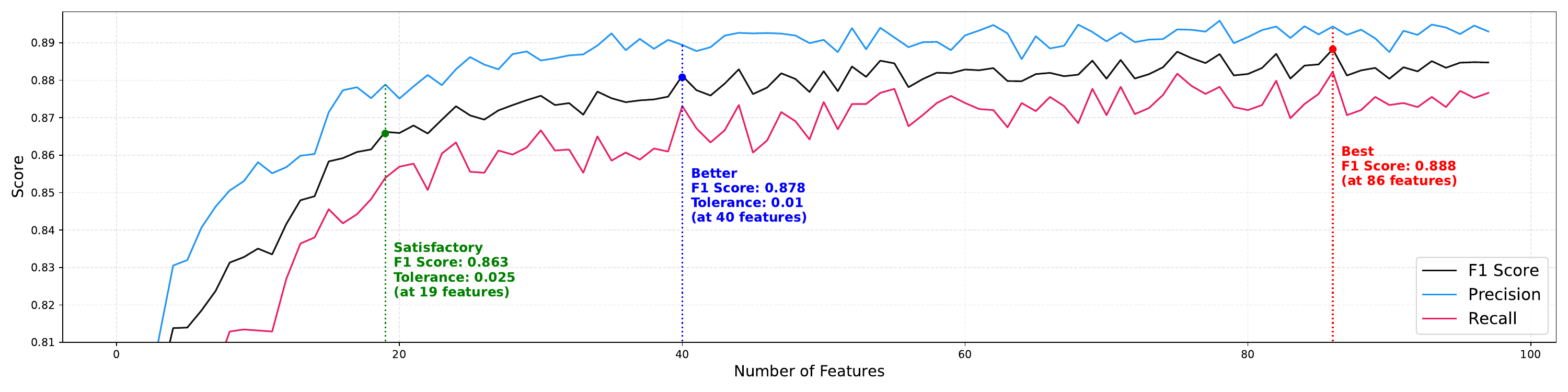}
        \caption{Control subjects and myocardial infarction patients classification with ascending counts for selected features.}
        \label{AFig6}
    \end{subfigure}
    \caption{Feature importance report and classification performance with selected features.}
\end{figure*}

\begin{figure*}[!t]
    \centering
    \begin{subfigure}[b]{\linewidth}
        \centering
        \includegraphics[width=\linewidth]{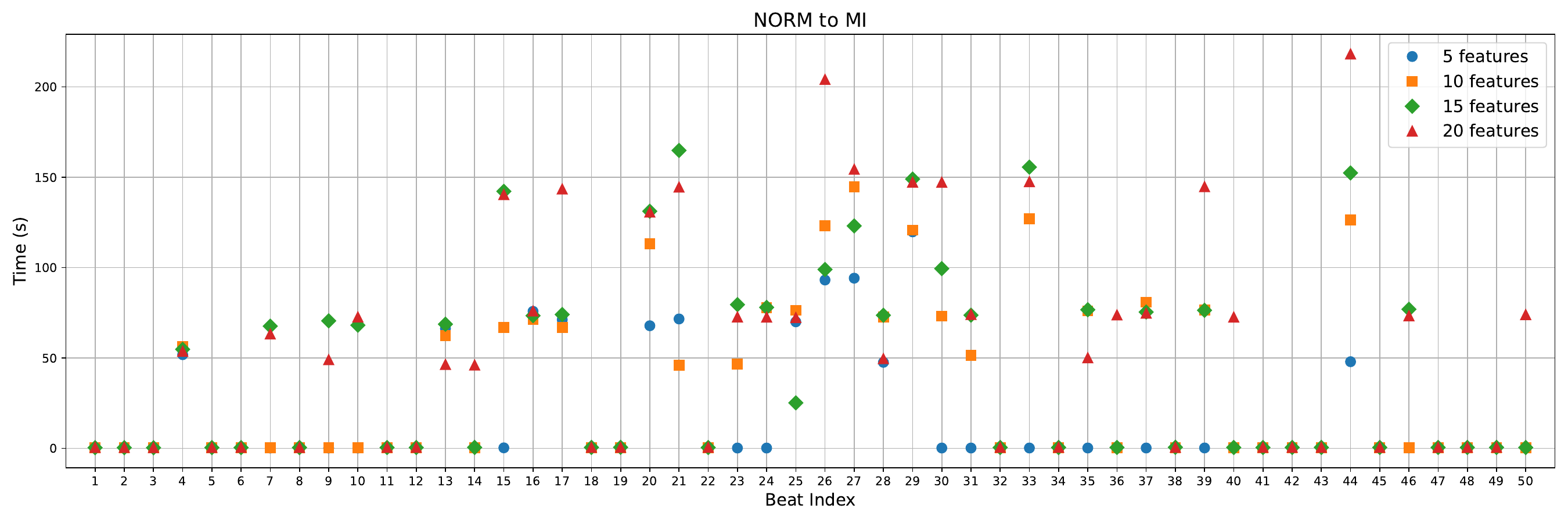}
        \caption{Required time to generate counterfactuals for NORM to MI.}
        \label{Afig:6}
    \end{subfigure}
    \hfill
    \begin{subfigure}[b]{\linewidth}
        \centering
        \includegraphics[width=\linewidth]{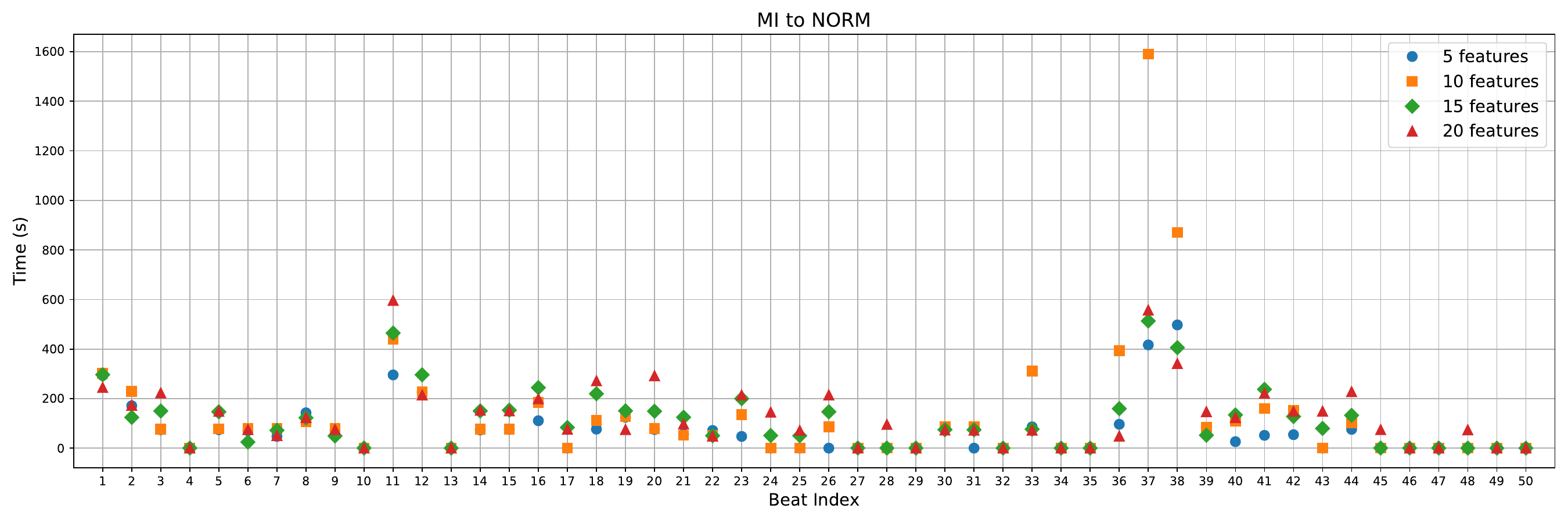}
        \caption{Required time to generate counterfactuals for MI to NORM.}
        \label{Afig:7}
    \end{subfigure}
    \caption{Comparison of time required to generate counterfactuals between NORM and MI, and vice versa.}
\end{figure*}

 To make the durations more conspicuous, you can see the elapsed times for sample 50 beats with 5, 10, 15, 20 feature counts in Figure \ref{Afig:6} and \ref{Afig:7}.

\begin{figure*}[tb!]
\centering
\includegraphics[width=0.82\linewidth]{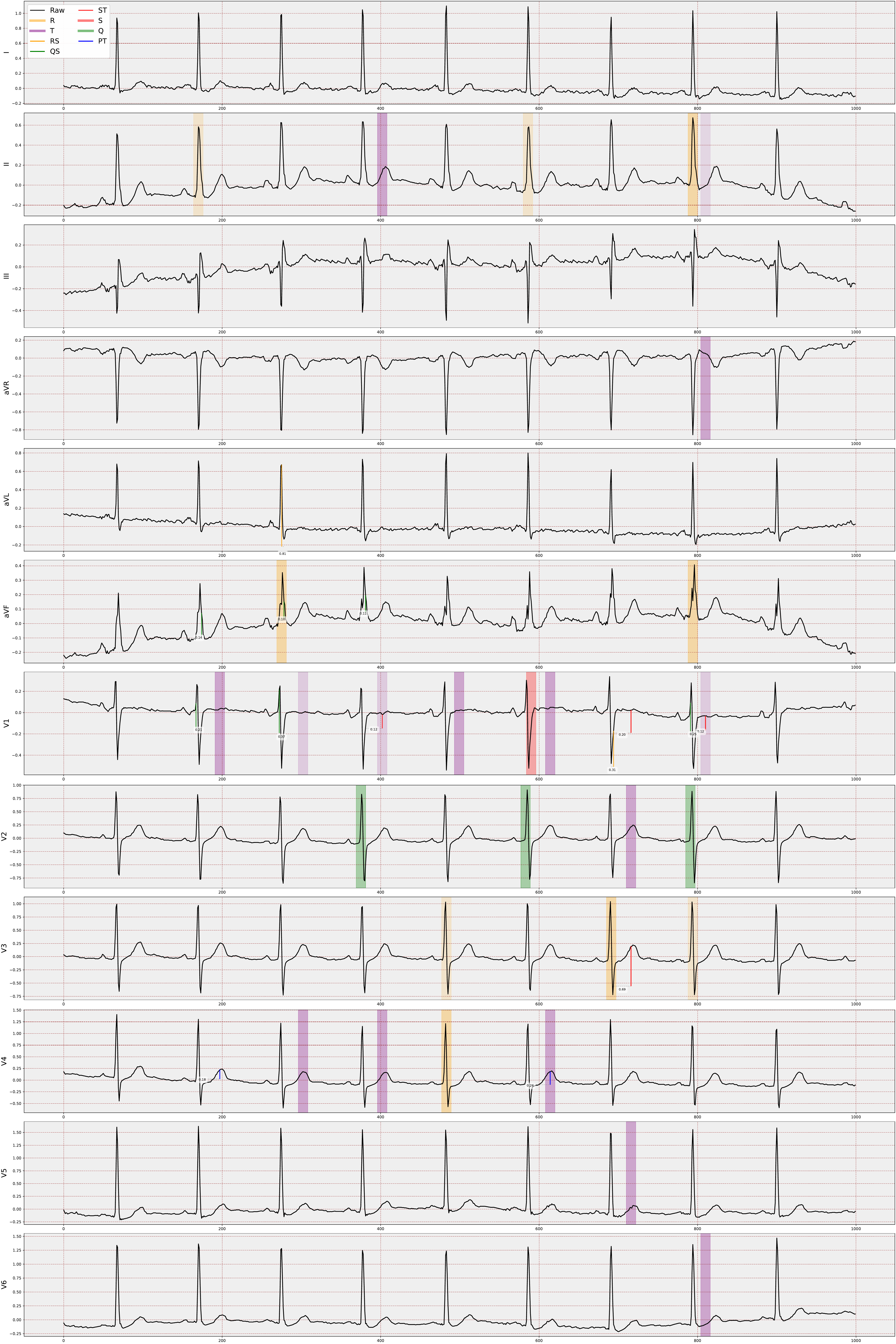}
\caption{Visualization of advanced features, which are derived from the amplitude disparities between two peaks, is depicted as lines superimposed on the ECG signal, with their respective distance values displayed alongside (id: 16773.0).}
\label{Afig:8}
\end{figure*}

\begin{figure*}[htb!]
\centering
\includegraphics[width=0.82\linewidth]{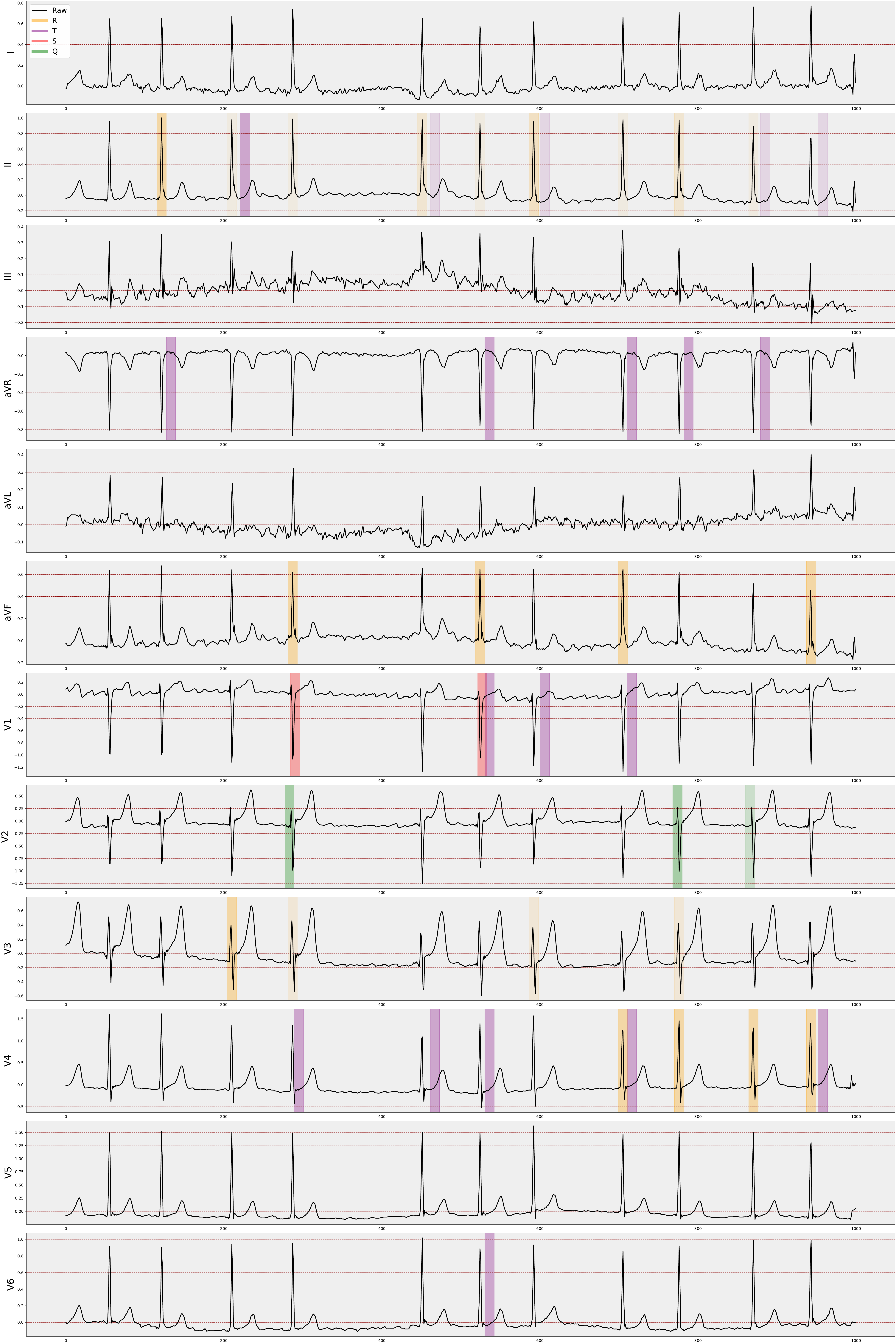}
\caption{Accurate Visualization (id: 1303.0): The specified ECG shows ST elevation in V1-V2-V3 derivations. The model correctly identified the loss of R progression in V3, deep S wave in V1, and ST-T changes in V1. In the PTB-XL dataset, this ECG was incorrectly evaluated as normal, our label is anterior MI.}
\label{Afig:9}
\end{figure*}

\begin{figure*}[htb!]
\centering
\includegraphics[width=0.83\linewidth]{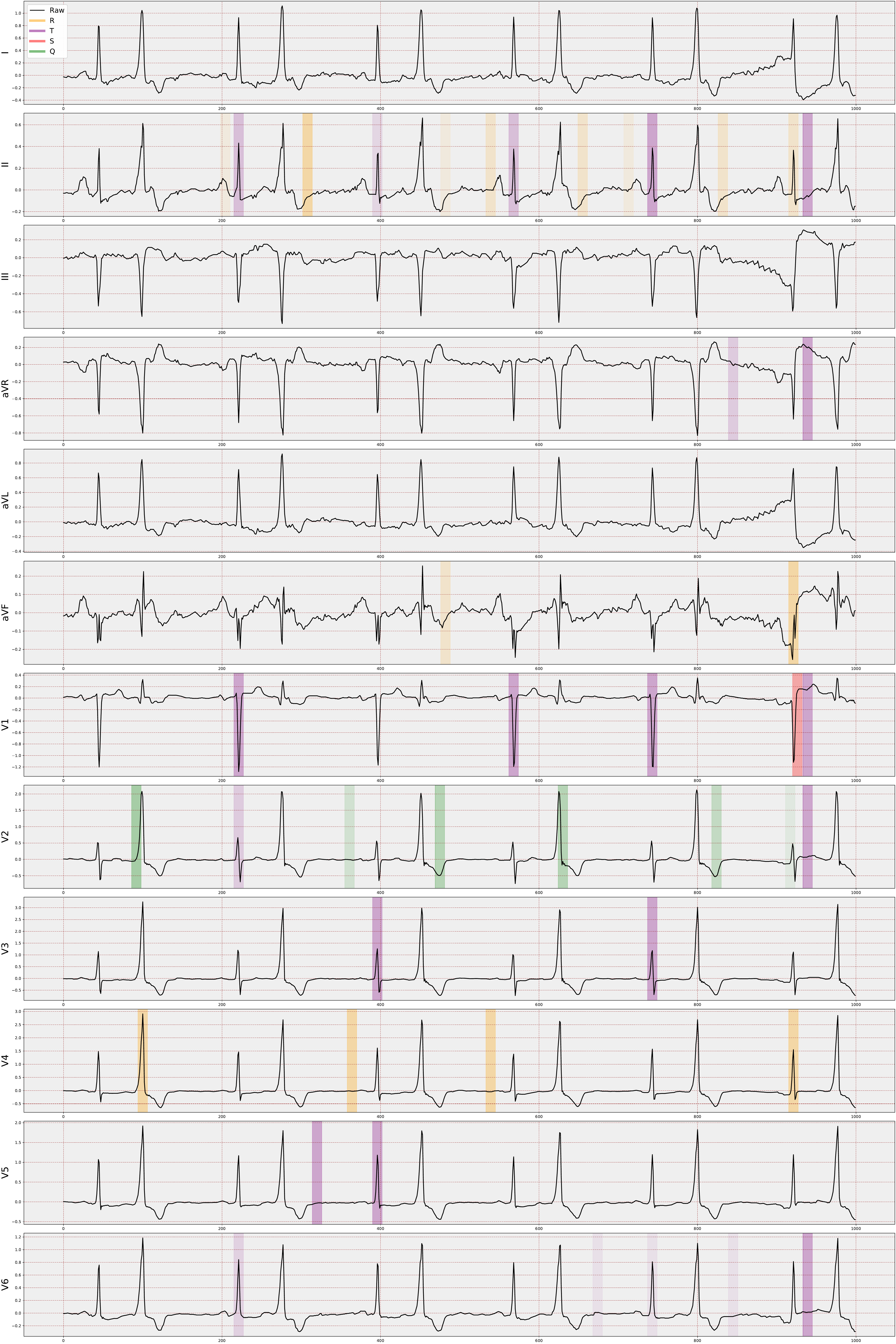}
\caption{False Detection Due to VES (id: 16448): In the specified ECG, the presence of bigeminy ventricular ectopic beats (VES) led the model to primarily evaluate these beats rather than the sinus beats (D2, V2, V4). This resulted in incorrect labeling by our method.}
\label{Afig:10}
\end{figure*}

\begin{figure*}[htb!]
\centering
\includegraphics[width=0.79\linewidth]{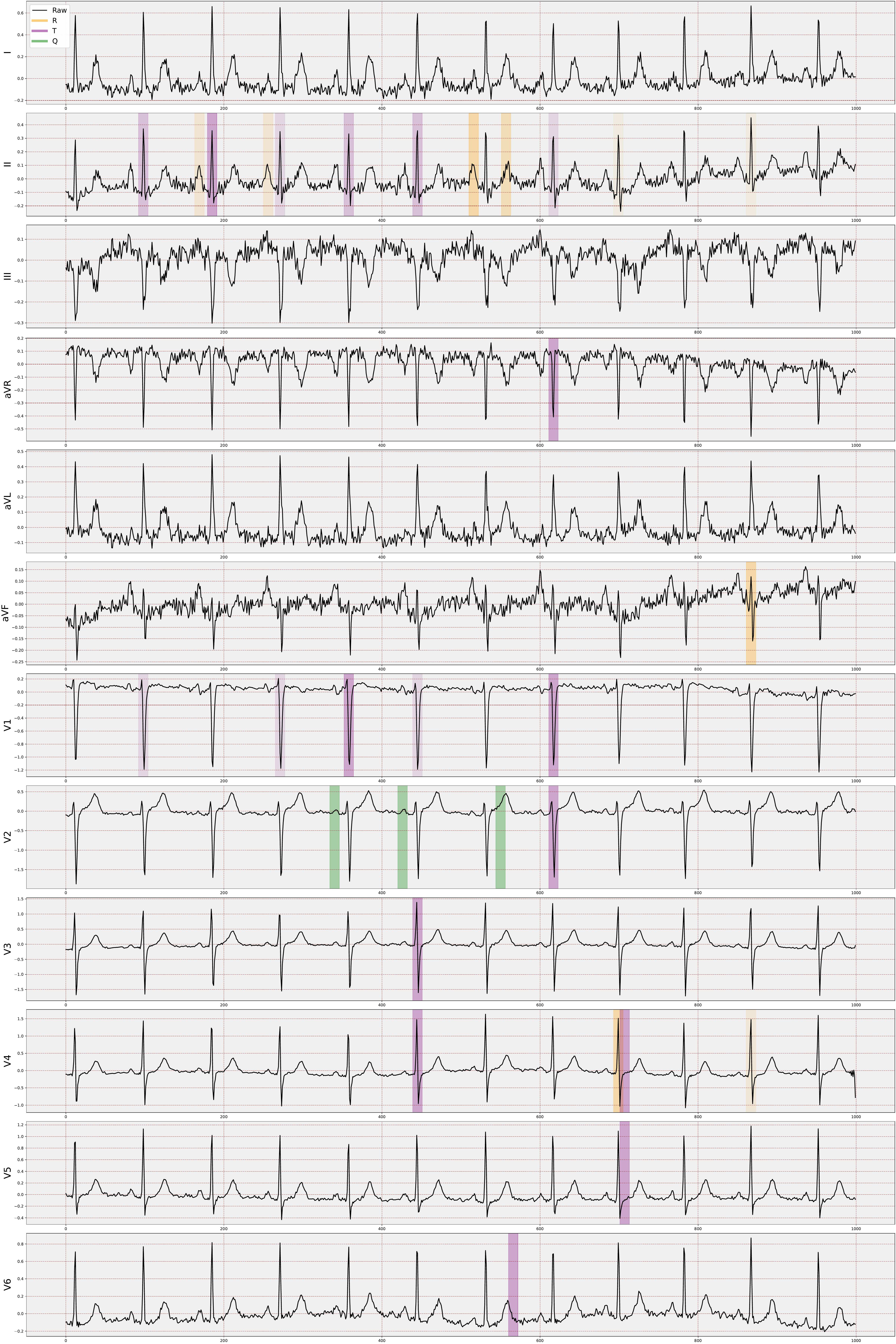}
\caption{False Detection Due to Artifact (id: 2261): In the specified ECG, artifacts observed, especially in the extremity derivations, could be attributed to possible muscle or neurological diseases. Consequently, our method exhibited a high error rate for this case, primarily due to the peak detection algorithm's inability to effectively manage the artifacts present in the signal. The PTB-XL dataset also contains instances where artifacts were mistakenly interpreted as potential inferior MIs, leading to incorrect diagnoses of inferior infarctions.}
\label{Afig:11}
\end{figure*}

\end{document}